\documentclass{tlp}
\usepackage{graphicx}
\usepackage{epic}
\usepackage{helvet}
\usepackage{courier}
\usepackage{amssymb}
\usepackage{amsmath}

\newcommand{\ignore}[1]{}
\newcommand{\finish}[1]{}
\newcommand{\skipit}[1]{{ #1 }}
 \newcommand{\nofootnote}[1]{}

\input{epsf}
\newtheorem{theorem}{Theorem}
\newtheorem{definition}{Definition}
\newtheorem{proposition}[theorem]{Proposition}
\newtheorem{corollary}[theorem]{Corollary}
\newtheorem{example}{Example}
\newtheorem{lemma}[theorem]{Lemma}

\newlength{\halftextwidth}
\newcommand{\SLIDE}{\mbox{\sc Slide}}
\newcommand{\REGULAR}{\mbox{\sc Regular}}
\newcommand{\CFG}{\mbox{\sc CFG}}
\newcommand{\GRAMMAR}{\mbox{\sc Grammar}}

\newcommand{\STRETCH}{\mbox{\sc Stretch}}

\newcommand{\AMONG}{\mbox{\sc Among}}

\newcommand{\SEQ}{\mbox{\sc Sequence}}

\newcommand{\GCC} {\mbox{\sc GCC}}
\newcommand{\SEQUENCE}{\mbox{\sc Sequence}}
\newcommand{\ALLDIFF}{\mbox{\sc AllDifferent}}

\newcommand{\SSUM} {\mbox{\sc SlidingSum}}
\newcommand{\RSSUM} {\mbox{\sc RelaxedSlidingSum}}
\newcommand{\SUM} {\mbox{\sc Sum}}
\newcommand{\CONTIG} {\mbox{\sc Contiguity}}
\newcommand{\PEAK} {\mbox{\sc Peak}}
\newcommand{\NOPEAK} {\mbox{\sc NoPeak}}
\newcommand{\HIGHESTPEAK} {\mbox{\sc HighestPeak}}
\newcommand{\CUMULATIVE} {\mbox{\sc Cumulative}}

\newcommand{\NVALUE}{\mbox{\sc NValue}}
\newcommand{\DIFFN}{\mbox{\sc Diffn}}

\newcommand{\DISJUNCTIVE}{\mbox{\sc Disjunctive}}

\newcommand{\DISJOINT}{\mbox{\sc Disjoint}}

\newcommand{\BINP}{\mbox{\sc BinPacking}}
\newcommand{\AVERAGE}{\mbox{\sc Average}}

\newcommand{\INTERDISTANCE}{\mbox{\sc InterDistance}}

\newcommand{\SPLASH}{\mbox{\sc Splash}}

\newcommand{\cA}{{\cal A}}
\newcommand{\cG}{{\cal G}}

\newcommand{\NN}{\ensuremath{\mathbb{N}}}
\newcommand{\ZZ}{\ensuremath{\mathbb{Z}}}

\newcommand{\decomp}{\mbox{\sc decomp}}

\begin{document}

\title{Contractibility for Open Global Constraints}

\author[M.J. Maher]
{Michael J. Maher \\
Reasoning Research Institute \\
Canberra, Australia    \\
E-mail: michael.maher@reasoning.org.au
}

\maketitle

\date{}

\begin{abstract}
Open forms of global constraints allow the addition of new variables to an argument during the execution of a constraint program.
Such forms are needed for difficult constraint programming problems
where problem construction and problem solving are interleaved,
and fit naturally within constraint logic programming.
However, in general, filtering that is sound for a global constraint can be unsound when
the constraint is open.
This paper provides a simple characterization, called contractibility, 
of the constraints where filtering remains sound when the constraint is open.
With this characterization we can easily determine whether a constraint has this property or not.
In the latter case, 
we can use it to derive a contractible approximation to the constraint.
We demonstrate this work on both hard and soft constraints.
In the process, we formulate two general classes of soft constraints.

Under consideration in Theory and Practice of Logic Programming (TPLP).
\end{abstract}


\section {Introduction}   \label{sect:intro}

Constraint Logic Programming (CLP) \cite{JM94}
provides the ability to add variables and constraints to a constraint store
during the course of an execution.
In this it is not alone:
linear and integer programming solvers and solvers presented as libraries
for an underlying programming language also allow
the introduction of new variables and constraints in an incremental way.
In some problems it is natural for the presence of some variables to be contingent
on the value of other variables.
This is true of configuration problems and scheduling problems that involve process-dependent activities \cite{condCSP,dynamic}.
More generally,
for difficult problems the intertwining of problem construction and problem solving
provides a way to manage the complexity of a problem,
and thus new variables and constraints may arise after solving has begun.
Thus CLP is particularly well-suited for such problems,
in contrast to compilation-based modelling languages such as MiniZinc \cite{MZ}
where all variables and constraints must be fixed at compilation time.

CLP also supports global constraints,
which have been an important part of the success of constraint programming.
However, most implementations of global constraints adopt a non-incremental approach:
the variables constrained by a global constraint are fixed when the constraint is imposed.
Thus the collection of variables they constrain is \emph{closed}, rather than \emph{open}.
This restricts the exploitation of incrementality that is available in CLP languages.
Delaying the imposition of a global constraint until all variables it might involve have been generated
can leave the filtering effect of the global constraint until too late
in the execution, resulting in a large search space.
Open global constraints 
remove this limitation by allowing variables to be added dynamically.

A major difficulty in implementing open constraints
is that a propagator for a closed constraint may be unsound
for the corresponding open constraint.
That is,
the propagator may make an inference that turns out
to be unjustified once the sequence of variables is extended.
In this paper we focus on the issue of identifying constraints for which
a closed propagator is sound as an open propagator.
These constraints have a simple characterization,
which we call \emph{contractibility},
and which allows us to easily determine whether a given constraint
has this property.
This characterization is also convenient for finding
the tightest contractible approximation of an uncontractible constraint,
which can be the basis for an open propagator of the constraint.
We illustrate our results with a wide variety of global constraints,
including both hard and soft constraints.

As part of our treatment of soft constraints we formulate two very general
classes of soft constraints based, respectively, on constraint decomposition and edit distance.
These classes unify and generalize several different proposals in the literature.
Using these formulations,
we introduce general results and techniques
for establishing that a constraint is contractible.
It turns out that finding a tightest contractible approximation is 
more difficult for soft constraints than for hard constraints.
In particular, while we can mathematically characterize the tightest approximation,
and define some pragmatic generic non-tight approximations,
we show that the tightest contractible approximation cannot always be represented in the edit-distance framework.

This paper is arranged as follows.
After some preliminaries in Section \ref{sect:prelim}
and a discussion of open constraints in Section \ref{sect:open},
we introduce contractibility in Section \ref{sect:cont}.
We show that it characterizes those constraints for which closed propagators remain sound
when the constraint is open,
and develop an algebra for constructing contractible constraints.
We conclude Section \ref{sect:cont} by
characterizing contractibility in language-theoretic terms,
and use that characterization to identify contractible constraints
(Section \ref{sect:class})
and tight approximations of uncontractible constraints
(Section \ref{sect:approx}).
We show that, with a tight approximation, a proposal of Bart\'{a}k for implementing
open uncontractible constraints achieves an appropriate consistency.
We then address the same issues for soft constraints
(Sections \ref{sect:softclass} and \ref{sect:softapprox}).

This paper incorporates results announced in \cite{open1,open2,open3,open4}.
It includes unpublished proofs, strengthened results, new results and some additional discussion.

\section{Background}   \label{sect:prelim}

The reader is assumed to have a basic knowledge of constraint programming, CSPs,
global constraints, and filtering,
as might be found in \cite{Dechter,CPhandbook,GCcatalog}.

For the purposes of this paper,
a global constraint is a relation over a single sequence of variables.
Other arguments of a constraint are considered parameters
and are assumed to be fixed before execution.
Throughout this paper,
a sequence of variables will be denoted, interchangeably, by $\vec{X}$ or $[X_1, \ldots, X_n]$.
We make no \emph{a priori} restriction on the variables
that may participate in the sequence
except that, in common with most work on global constraints,
we assume that no variable appears more than once in a single constraint.

There are some specific global constraints that we define for completeness.
These and other global constraints are discussed more completely in \cite{GCcatalog}
and the references therein.
As with variables, a sequence of values $v_i$ is expressed by $\vec{v}$.
The constraint $\ALLDIFF([X_1, \ldots, X_n])$ \cite{regin}
states that the variables $X_1, \ldots, X_n$ take distinct values.
The global cardinality constraint \linebreak
$\GCC(\vec{v}, \vec{l}, \vec{u}, [X_1, \ldots, X_n])$ \cite{regin96}
states that, for every $i$,
the value $v_i$ occurs between $l_i$ and $u_i$ times in the list of variables.
The constraint $\NVALUE([X_1, \ldots, X_n], N)$ \cite{nvalue} states that 
there are exactly $N$ distinct values in $X_1, \ldots, X_n$.
The constraint $\REGULAR(\cA, [X_1, \ldots, X_n])$ \cite{regular} states that
the value of the list of variables, when considered as a word,
is accepted by the automaton $\cA$.
Similarly, the constraint $\CFG(\cG, [X_1, \ldots, X_n])$  \cite{QW,Sellmann}
(called $\GRAMMAR$ in \cite{QW})
states that the value of the list of variables, when considered as a word,
is generated by the context-free grammar $\cG$.

The constraint $\SEQUENCE(l, u, k, [X_1, \ldots, X_n], \vec{v})$ \cite{BC} states that
any consecutive sequence of $k$ variables $X_j, \ldots, X_{j+k-1}$
contains between $l$ and $u$ occurrences of values from $\vec{v}$.
The constraint $\SSUM(l, u, k, [X_1, \ldots, X_n])$ \cite{cardpath} states that
the sum of any consecutive sequence of $k$ variables lies between $l$ and $u$.
The constraint $\CONTIG([X_1, \ldots, X_n])$ \cite{contig} states that
the variables $X_i$ take values from $\{0, 1\}$ and the variables
taking the value 1 are consecutive.
The lexicographical ordering constraint $[X_1, \ldots, X_n] \leq_{lex} [Z_1, \ldots, Z_n]$ \cite{lex}
states that the sequence of $X$ variables is lexicographically less than or equal to
the sequence of $Z$ variables, where we assume some ordering on the underlying values.
The precedence constraint $s \prec_{\vec{X}} t$ \cite{precedence} states that
if $t$ appears in the sequence $\vec{X}$ then $s$ appears at a lower index.

For some constraints, like $\ALLDIFF$, $\GCC$ and $\NVALUE$, the order of variables is 
immaterial to the semantics of the constraint.
We say a constraint $C$ is 
\emph{order-free}
if
\[
C([X_1, \ldots, X_n]) \leftrightarrow C([X_{\pi(1)}, \ldots, X_{\pi(n)}])
\]
for every permutation $\pi$ of $1..n$.
The other constraints mentioned above are not order-free.

We assume that the argument $\vec{X}$ of a use of a global constraint has a static type $T$ that
assigns, for every position $i$, a set of values.
Thus every variable $X$ in $\vec{X}$ has a static type $T(X)$ of values that it may take.
We will also view $T$ as a unary predicate on the variables of $\vec{X}$,
where $T(X)$ is true iff $X$ takes a value from its static type.
In addition,
generally, each variable has an associated set $S \subseteq T(X)$ of values, called its domain.
We will view this simultaneously as: 
a function $D : \vec{X} \rightarrow 2^{Values}$ where $D(X) = S$ and $Values = \bigcup_{X in \vec{X}} T(X)$,
a unary relation $D(X)$ which is satisfied only when the value of $X$ is some $s \in S$,
and the pointwise extension of $D$ to sequences of variables.

We formalize the semantics of a global constraint $C$ as a formal language $L_C$.
A word $d_1 d_2 \ldots d_n$ appears in $L_C$ iff
the constraint $C([X_1, X_2, \ldots, X_n])$ has a solution
$X_1 = d_1, \ldots X_n = d_n$.
Thus, for example,
the semantics of $\ALLDIFF$ is
$\{ a_1 \ldots a_n ~|~ \forall i, j ~ i \neq j \rightarrow a_i \neq a_j, n \in \NN \}$ and
the semantics of $\REGULAR(\cA, \vec{X})$ is $L(\cA)$, the language accepted by $\cA$.
When it is convenient, we will describe languages with Kleene regular expressions \cite{HU}.
For a given use of a constraint $C(\vec{X})$, we write $T(\vec{X})$ for the language defined by the static type of $C(\vec{X})$.

The following definitions will be important later.
Let $P(L) = \{ w ~|~ \exists u ~ wu \in L \}$ denote the set of prefixes of a language $L$,
called the \emph{prefix-closure} of $L$.
We say $L$ is \emph{prefix-closed} if $P(L) = L$.
We say two languages $L$ and $L'$ are \emph{prefix-equivalent} if $P(L) = P(L')$.

\section {Open Constraints}  \label{sect:open}

There are many problems that are dynamic in nature
but to which we would like to apply constraint techniques.
\cite{Bartak99} describes a class of complex processing environments
where there may be alternative processing routes, different production formulas
and alternative raw materials.
In addition to the core products of the processes, there may be by-products and co-products
which require additional processing.
Some instances of products may be re-processed or recycled.
Because of storage limits and/or a necessity to work with the instances while they are still in an amenable state,
such instances might need to be re-processed or recycled promptly.
In such environments, process scheduling must be dynamic:
additional tasks may arise from re-processing, and additional raw materials may arise.

Many production processing environments have these characteristics.
Consider, for example, sugar cane processing.
Juice is extracted from the sugar cane and clarified before it is refined.
Refining involves repeated crystallization and centrifuging processes,
with molasses produced as a by-product.
Usually three repetitions of these processes are performed but,
through natural variation of the raw materials,
an additional repetition may be needed.
Such a need can be identified through monitoring the refinement process.

Now consider a constraint-based approach to the problem of the on-going scheduling of these processes.
We might use a $\CUMULATIVE$ constraint to express the limited availability of centrifuges.
When a batch requires an additional repetition, a new task must be added to that constraint
and additional constraints concerning the task must be added to the problem.
Thus we require that $\CUMULATIVE$ be an \emph{open} constraint --
able to accept additional tasks.

Open constraints pre-suppose the existence of a meta-program
that can 
impose constraints, 
close an open constraint,
add variables to an open constraint,
(possibly) create new variables,
and interact with the execution of the constraint system,
possibly controlling it.
In this paper we will abstract away the details of the meta-program
so that we can focus on the open constraints.
We assume that the collection of variables forms a sequence,
to which variables may be added at the right-hand end only.\footnote{
There is a brief discussion of the effect of alternatives in Section \ref{sect:disc}.
}
The scope of constraints changes during the execution,
and we refer to the state of the constraint at some point in the execution
as an \emph{occurrence} of the constraint.
In open global constraints $C$ the length of the sequence of variables varies
and consequently the semantics in terms of the language $L_C$ is particularly appropriate.

There are three models of open constraint that have been proposed.\footnote{
The terminology ``open constraint satisfaction problem'' was introduced
by 
\cite{openCS,openCP}.
However, that use refers to problems in which the set of variables is closed
but the domains are open,
that is, extra values can be added to variable domains.
That work is not technically related to ``open constraints'' as used in this paper,
but it shares with this paper an interest in constraint problems that may change over time.
}
\cite{dynamic} first formulated this issue and described a straightforward model:
the constraint involves a sequence of variables to which variables may be added.
Thus the arity and type of the constraint are unchanged,
whether the constraint is open or closed.
\cite{dynamic} outlined
a generic implementation technique to make open versions for the class of 
\emph{monotonic} global constraints.
Bart\'{a}k focussed on a specific implementation of the open $\ALLDIFF$ constraint.
This is an order-free constraint,
and details of the model, such as where variables are added to the sequence,
are left unspecified.
The remaining models extend this model by incorporating more details about the possible
extension of the sequence; for these models the constraint has a different arity or type.

The model of 
\cite{open}
only applies to order-free constraints expressed in the form $C(S)$.
It uses a set variable $S$ describing a \emph{set} of object variables, rather than a sequence, 
to represent the collection of variables in the constraint.\footnote{
A set variable $S$ ranges over sets and 
is constrained by two fixed finite sets $L$ and $U$ which are a lower and upper bound
on the value of the variable: $L \subseteq S \subseteq U$.  
See \cite{conjunto}.
}
The lower bound of $S$ is the set of variables that are committed to appear in the constraint;
the upper bound is the set of variables that are permitted to appear in the constraint.
Thus there is a finite set of variables that might appear in the constraint,
and these are fixed in advance.
The authors refer to the constraint as  open ``in a closed world''
since the set of variables that might be added to the constraint is closed.
The model makes elegant use of existing implementations of
set variables and their associated bounds.
However, the
use of a constraint in this model requires knowing all the variables that might appear
before imposing the constraint.
As a result, it cannot deal well with contingent variables.
They create a similar problem to the one faced by closed constraints:
the constraint may be imposed late in the execution,
creating a larger search space.

The third model \cite{open2}
is, in some ways, intermediate between
that of \cite{dynamic} 
and  \cite{open}.
Under this model,
a constraint $C(\vec{X}, N)$ acts on both a sequence of variables $\vec{X}$ and
an integer variable $N$ representing the length of the sequence once it is closed.
Variables can only be added at the right-hand end of the sequence.
This is a more detailed model than Bart\'{a}k's.
In one sense, this model is an abstraction of the model of  \cite{open}:
if $N$ is subject only to lower and upper bounds,
then the bounds on $N$ correspond to the cardinalities of the bounds of $S$.
It does not have the weakness of that model that the variables that might appear
are fixed in advance.
On the other hand, the van Hoeve-Regin model has more information about how
$\vec{X}$ might be extended, and so might be able to perform stronger propagation.

The model we employ here is Bart\'{a}k's model where we specify that
variables may be added only at the right-hand end of the sequence.  
It is equivalent to a weak form of the model of \cite{open2}
where there are no restrictions on $N$.
However, the notion of contractibility, to be introduced in the next section,
is relevant for other models of open constraints.
Some results are given in \cite{open2} for the model treated there.
We will assume that the only operations that can be applied to an open constraint
are adding a variable and designating the constraint closed, so that no more variables may be added.

Constraint programming with open constraints is a special case of 
dynamic CSPs in the broad sense described in \cite{DCSP}.
Work on dynamic CSPs has focussed on the addition and retraction of constraints
\cite{VHLP,Bess91,retraction,Debetal03}.
It does not directly address the addition of variables to a constraint,
although that can be viewed as a combined retraction and addition of constraints.
See \cite{dynCSP} for a survey on dynamic constraint solving.
Work on conditional CSPs, initiated in \cite{condCSP}, addresses contingent variables
by explicitly embedding the contingent nature within a CSP,
but that work does not address the addition of variables to constraints.

Other forms of dynamism have been addressed in the context of constraints
by allowing variable domains to be initially incomplete and expand over time \cite{openCP,incomplete_domains},
or by formulating constraints over a stream of values \cite{CPstreams}.
That work is not technically related to the work in this paper.

We take \emph{filtering} or \emph{propagation} to refer to any algorithm $f$ that reduces domains, 
that is, $\forall X ~ f(D)(X) \subseteq D(X)$.
A filtering algorithm $f$ for a constraint $C$ is \emph{sound} if
every solution of $C$ in $D$ also appears in $f(D)$.
Some filtering algorithms are characterized by consistency conditions.
For closed constraints, the strongest filtering/consistency condition that addresses
each constraint separately is domain consistency.
A closed constraint $C(X_1, X_2, \ldots, X_n)$ is \emph{domain consistent} if
for every $i$ where $1 \leq i \leq n$ and every $d \in D(X_i)$
there is a word $d_1 \ldots d_n$ in $L_C$ such that
$d_i = d$ and
$d_j \in D(X_j)$ for $j = 1,\ldots,n$.

Because some of the variables in an open constraint will be unspecified during part of the execution,
we need to adapt the definition of consistency.
The following is an appropriate form of domain consistency  for Bart\'{a}k's model.
\begin{definition}
Given a domain $D$,
an occurrence of a constraint $C(\vec{X})$ is
\emph{open D-consistent} if,
for every $X_i \in \vec{X}$ and every $d \in D(X_i)$,
there is a word $d_1 \ldots d_m$ in $L_C$ such that
$d_i = d$,
$|\vec{X}| \leq m$,
and
$d_j \in D(X_j)$ for $j = 1,\ldots,|\vec{X}|$.
\end{definition}

When $C$ is closed, the only words of interest in $L_C$ are those of length $|\vec{X}|$.
In that case open D-consistency reduces to domain consistency.

\section {Contractibility}  \label{sect:cont}

We want to extend a constraint $C([X_1, \ldots, X_n])$ with an extra variable $Y$ to $C([X_1, \ldots, X_n, Y])$.
We would like to do filtering on the smaller constraint without knowing whether it
will be extended to $Y$, or further, and without creating a choicepoint.
When we can do this, we have a kind of monotonicity property of $C$.

\begin{definition}   \label{defn:cont}
We say a constraint $C([X_1, \ldots, X_n])$ is \emph{contractible} if,
there is a number $m$ such that for all $n \geq m$ we have
\[
C([X_1, \ldots, X_n, Y]) \rightarrow  C([X_1, \ldots, X_n])
\]
The least such $m$ is called the \emph{contractibility threshold}.

For this paper we consider only constraints with a contractibility threshold of 0.
\end{definition}

Thus $C$ is contractible iff  every solution of $C([X_1, \ldots, X_n, Y])$,
when restricted to $X_1, \ldots, X_n$ where $m \leq n$,
is a solution of $C([X_1, \ldots, X_n])$.
The property is akin to the ``optimal substructure'' property
that is a pre-requisite for the use of dynamic programming in optimization problems
\cite{CLRS}
which requires that optimal solutions of a problem also solve subproblems optimally.
Here it is only satisfiability, and not optimality, that is involved.

It follows that any sound form of filtering
(such as arc consistency or bounds consistency)
on a contractible constraint $C([X_1, \ldots, X_k])$ is safe
in the sense that any values deleted from domains in that process could also be deleted
while filtering on $C([X_1, \ldots, X_n])$ for any $n \geq k$.
Recall that we use $\vec{X}$ and $[X_1, \ldots, X_n]$ interchangeably.

\begin{proposition}   \label{prop:filter}
Let $C$ be a contractible constraint.
Suppose a sound filtering algorithm for $C([X_1, \ldots, X_n])$
reduces the domain $D$ for $\vec{X}$ to $D'$.
Then
\[
D(\vec{X}) \wedge C([X_1, \ldots, X_n, Y]) \leftrightarrow D'(\vec{X}) \wedge C([X_1, \ldots, X_n, Y])
\]
Furthermore,
if this property holds for all domains and all sound filterings
then $C$ must be contractible.
\end{proposition}
\skipit{
\begin{proof}
Let $\sigma$ be a solution of $D(\vec{X}) \wedge C([X_1, \ldots, X_n, Y])$.
By contractibility of $C$, $\sigma$ satisfies $C([X_1, \ldots, X_n])$.
By the soundness of the filtering, $\sigma$ satisfies $D'(\vec{X})$.
Hence, $\sigma$ satisfies $D'(\vec{X}) \wedge C([X_1, \ldots, X_n, Y])$.
Since $\sigma$ is an arbitrary solution,
\[
D(\vec{X}) \wedge C([X_1, \ldots, X_n, Y]) \rightarrow D'(\vec{X}) \wedge C([X_1, \ldots, X_n, Y])
\]
Since $D'$ results from filtering $D$, 
$D'(\vec{X}) \rightarrow D(\vec{X})$
and hence the reverse direction also holds.

Now, 
suppose this property holds for all sound filterings $D \leadsto D'$
but $C$ is not contractible.
Because $C$ is not contractible,
there must be a number $n$ and a valuation $\sigma$ 
that satisfies $C([X_1, \ldots, X_n, Y])$ but not $C([X_1, \ldots, X_n])$.
Let $D$ be the domain that defines $\sigma$ and $D'$ be the empty (unsatisfiable) domain.
Then the reduction of $D$ to $D'$ is sound for $C([X_1, \ldots, X_n])$ and so,
by the previous supposition
\[
D(\vec{X}) \wedge C([X_1, \ldots, X_n, Y]) \rightarrow D'(\vec{X}) \wedge C([X_1, \ldots, X_n, Y])
\]
However,
$D(\vec{X}) \wedge C([X_1, \ldots, X_n, Y])$ is satisfiable by $\sigma$, while $D'(\vec{X})$
is unsatisfiable, which contradicts this statement.
This contradiction shows that $C$ must be contractible.
\end{proof}
}

Consequently, for contractible constraints, filtering
does not need to be undone if the list is lengthened.
That is, algorithms for filtering a closed contractible constraint are valid also for 
the corresponding open constraint.

Conversely, any constraint that is not contractible might need to undo the effects of filtering
if the list is lengthened.
If $\sigma$ is a solution of $C([X_1, \ldots, X_n, Y])$,
but not of $C([X_1, \ldots, X_n])$
then propagation on $C([X_1, \ldots, X_n])$ might eliminate $\sigma$.
For example,
a constraint
$\sum_{i} X_i = 5$
would propagate $X_1 = 5$ if the sequence $\vec{X}$ contains just one variable,
thus eliminating solutions such as $X_1 = 2, X_2 = 3$.
When the second variable is added,
all propagation that is a consequence of the inference $X_1 = 5$
must be undone.

The second part of this proposition shows that
contractibility exactly characterizes the guarantee that
closed filtering is safe for open constraints.
That is, it is exactly the contractible constraints for which it is always sound to interleave
closed filtering and addition of new variables.

Furthermore, the proof of the second part requires very little of the filtering algorithm.
Hence, whether we maintain arc consistency or weaker consistencies like
bounds consistency or forward checking,
contractibility is necessary to soundly interleave closed filtering and the addition of new variables.

We say a domain $D$ \emph{defines an assignment} if $\forall X ~ |D(X)| = 1$;
in that case the assignment maps each $X$ to the element of $D(X)$.
We say filtering performs \emph{complete checking}
if, whenever $D$ defines an assignment, the result of filtering with a constraint $C$
is $D$ iff the assignment satisfies $C$.
Complete checking can be considered a minimal requirement for filtering methods \cite{ST}.
Any filtering method that satisfies this minimal requirement requires contractibility
to guarantee that closed filtering is sound for an open constraint.

\begin{corollary}   \label{cor:cont}
Let $C$ be a constraint,
and consider a sound filtering method that performs complete checking.
It is always sound to interleave filtering and the addition of new variables
iff $C$ is contractible.
\end{corollary}

The notion of contractibility is a variation of Bart\'{a}k's monotonicity \cite{dynamic}
where 
we do not explicitly discuss variable domains.
Before proceeding, we make this claim precise.
We formulate Bart\'{a}k's monotonicity as follows.

\begin{definition}
Let $D$ be a domain.
We say a constraint $C$ is \emph{monotonic} with respect to $D$ if,
for any pair of disjoint sequences of variables $\vec{X}$ and $\vec{Y}$
\[
\{ \vec{X} ~|~  C(\vec{X} \vec{Y}) \wedge D(\vec{X}) \wedge D(\vec{Y}) \}
 \subseteq 
\{ \vec{X} ~|~  C(\vec{X}) \wedge D(\vec{X}) \}
\]
\end{definition}

Contractibility differs from monotonicity in that
the definition is based entirely on the constraint,
independent of the domains of variables.
Hence it is not tied to domain-based reasoning;
it is equally compatible with the more general framework of \cite{synth}.
On the other hand, monotonicity is more flexible in reasoning about
constraints that are only ``partly contractible''.
The close relationship between monotonicity and contractibility is clear.

\begin{proposition}   \label{prop:mono}
If $C$ is contractible then 
for any domain $D$, $C$ is monotonic with respect to $D$.
Conversely,
if $C$ is monotonic with respect to every domain $D$ then $C$ is contractible.
\end{proposition}
\skipit{
\begin{proof}
By repeated application of the definition of contractibility, we have that\linebreak
$C([X_1, \ldots, X_n, \vec{Y}]) \rightarrow  C([X_1, \ldots, X_n])$.
It follows immediately that $C$ is monotonic  with respect to any particular $D$.

In the reverse direction, any valuation for $\vec{X}\vec{Y}$ can be represented by a domain $D$
where each $D(X_i)$ and $D(Y_i)$ is a singleton.
Then monotonicity with respect to $D$ implies
$C(\vec{X} \vec{Y})  \rightarrow  C(\vec{X})$ under that valuation.
If $C$ is monotonic with respect to every domain $D$ then
$C(\vec{X} \vec{Y})  \rightarrow  C(\vec{X})$ holds under every valuation.
That is, $C$ is contractible.
\end{proof}
}

We now turn to ways a constraint can be constructed to ensure it is contractible.
As a trivial case,
a constraint $C$ of fixed arity $k$, when applied to a sequence of variables $\vec{X}$, is assumed to
be applied only to the initial segment $X_1, \ldots, X_k$, or not at all if $\vec{X}$ is shorter than $k$.
With this definition, $C$ is contractible.

The $\SLIDE_j$ meta-constraint \cite{slide}
can be used to define several constraints on a sequence of variables.
We use a variant of $\SLIDE_j$ that starts applying the constraint at the $p^{\mbox{th}}$ position,
rather than the first.
$\SLIDE^p_j(C, \vec{X})$ holds iff
$C(X_{ij+p}, \ldots, X_{ij+p+k-1})$ holds for $i = 0, 1, \ldots, \lfloor \frac{n-p-k+1}{j} \rfloor$,
where $C$ has arity $k$.
$\SLIDE_j$ is equal to $\SLIDE^1_j$.

Constraints defined directly with $\SLIDE_j^p$ are contractible.
\begin{proposition}   \label{prop:slidepj}
Any constraint $C$ defined by the $\SLIDE^p_j$ meta-constraint as
$C(\vec{X}) \leftrightarrow \SLIDE^p_j(C', \vec{X})$,
for some fixed arity constraint $C'$, is contractible.
\end{proposition}
\skipit{
\begin{proof}
Let $k$ be the arity of $C'$.
The relationship between  $C([X_1, \ldots, X_n, Y])$ and $C([X_1, \ldots, X_n])$
divides into cases, using the definition of $C$.
If $n-p-k+2$ is non-negative and divisible by $j$ then
\[
C([X_1, \ldots, X_n, Y]) \leftrightarrow
C([X_1, \ldots, X_n]) \wedge C'([X_{n-k+2}, \ldots, X_n, Y])
\]
If $n-p-k+2$ is negative or not divisible by $j$ 
then there is no additional application of $C'$ and
\[
C([X_1, \ldots, X_n, Y]) \leftrightarrow
C([X_1, \ldots, X_n])
\]
Thus, in both cases,
$C([X_1, \ldots, X_n, Y]) \rightarrow  C([X_1, \ldots, X_n])$
and hence $C$ is contractible.
\end{proof}
}

Since the $\SEQUENCE$ and $\SSUM$ constraints can each be defined as $\SLIDE_1(C', \vec{X})$,
for appropriate constraint $C'$, 
it follows that they are both contractible.

For order-free constraints 
we can define a meta-constraint analogous to $\SLIDE$,
which we will call $\SPLASH$.
Like $\SLIDE$, it takes a fixed arity constraint $C'$ and a sequence of variables $\vec{X}$ as arguments. 
Let $C'$ have arity $k$, and $\vec{X}$ have length $n$, and let
$S_k(\vec{X}) = \{ [X_{i_1}, \ldots, X_{i_k}] ~|~ i_j < i_{j+1} \mbox{ for } j=1,...,k-1 \}$
be the set of subsequences of $\vec{X}$ of length $k$.
Then we define
$\SPLASH(C', \vec{X}) \leftrightarrow \bigwedge_{\vec{Y} \in S_k(\vec{X})} C'(\vec{Y})$.
$\SPLASH(C', \vec{X})$
applies $C'$ to every subsequence of $\vec{X}$ of length $k$.
For example, we can define $\ALLDIFF(\vec{X})$ as $\SPLASH(\neq, \vec{X})$
and $\INTERDISTANCE(\vec{X})$ as $\SPLASH(C', \vec{X})$
where $C'(Z_1, Z_2)$ $ \leftrightarrow |Z_1 - Z_2| \geq p$.
Thus, by the following proposition, $\ALLDIFF$ and $\INTERDISTANCE$ are contractible.

\begin{proposition}   \label{prop:splash}
Any constraint $C$ defined by the $\SPLASH$ meta-constraint as
$C(\vec{X}) \leftrightarrow \SPLASH(C', \vec{X})$,
for some fixed arity constraint $C'$, is contractible.
\end{proposition}
\skipit{
\begin{proof}
Let $k$ be the arity of $C'$.
It is straightforward to see that
\[
C([X_1, \ldots, X_n, Y]) \leftrightarrow
C([X_1, \ldots, X_n]) \wedge \bigwedge_{\vec{Z} \in S_{k-1}(\vec{X})}C'([Z_{1}, \ldots, Z_{k-1}, Y])
\]
It follows immediately from the definition that $C$ is contractible.
\end{proof}
}

Once we have some contractible constraints, there are many ways to build other contractible constraints,
as the following proposition demonstrates.
These are expressed as logic operators, but they can also be viewed as operators on
formal languages:
$\wedge$ and $\vee$ are intersection and union of languages, negation is complement,
existential quantification projects out a variable,
and
universal quantification retains words that appear for all values of the relevant variable.

\begin{proposition}   \label{prop:conserve}
Let $C_1(\vec{X})$ and $C_2(\vec{X})$ be contractible constraints on the same sequence of variables.
Let $C(X_1, \ldots, X_k)$ be a constraint of fixed arity.
Then
\begin{itemize}
\item $C$ is contractible
\item $C_1 \wedge C_2$ is contractible
\item $C_1 \vee C_2$ is contractible
\item $\exists X_i ~ C_1$ is contractible
\item $\forall X_i ~ C_1$ is contractible
\end{itemize}
where $X_i$ is a variable in $\vec{X}$.
\end{proposition}
\skipit{
\begin{proof}
We can view $C$ as a constraint $C'$ on the sequence $\vec{X}$
where $C'([X_1, \ldots, X_n]) \leftrightarrow true$ if $n < k$
and $C'([X_1, \ldots, X_n]) \leftrightarrow C(X_1, \ldots, X_k)$
if $n \geq k$.
Note that $C(X_1, \ldots, X_k)$ $\rightarrow true$
and hence
$C'([X_1, \ldots, X_{k-1}, Y]) \rightarrow  C'([X_1, \ldots, X_{k-1}])$.
When $n \neq k-1$ we clearly have
$C'([X_1, \ldots, X_n, Y]) \leftrightarrow  C'([X_1, \ldots, X_n])$.

Suppose
$C_i([X_1, \ldots, X_n, Y]) \rightarrow  C_i([X_1, \ldots, X_n])$
for $i=1,2$.
Then, by propositional logic,
\[\bigwedge_i C_i([X_1, \ldots, X_n, Y]) \rightarrow  \bigwedge_i C_i([X_1, \ldots, X_n])\]
and 
\[\bigvee_i C_i([X_1, \ldots, X_n, Y]) \rightarrow  \bigvee_i C_i([X_1, \ldots, X_n])\]
Similarly, 
using standard arguments,
for any $i$ we can conclude
\[\forall X_i ~ C_1([X_1, \ldots, X_n, Y]) \rightarrow  \forall X_i ~ C_1([X_1, \ldots, X_n])\]
and
\[\exists X_i ~ C_1([X_1, \ldots, X_n, Y]) \rightarrow  \exists X_i ~ C_1([X_1, \ldots, X_n])\]
\end{proof}
}

In general, the negation of a contractible constraint and
implication between two contractible constraints are not contractible.
See Example \ref{ex:negimp}, later.

The previous results give us an algebra for constructing complex contractible constraints,
and can be used to demonstrate that some existing constraints are contractible.
For example,
$\CONTIG$ is implemented in \cite{contig} essentially as
$$\exists \vec{L}, \vec{R} ~ SLIDE^2_3(C', [L_1, X_1, R_1, L_2, \ldots, X_n, R_n])$$
where $C'$ has arity 7.
Similarly, $(\vec{X} \leq_{lex} \vec{Y})$ is encoded in \cite{slide} essentially as
$$\exists \vec{B} ~ SLIDE_3(C', [B_1, X_1, Y_1, B_2, \ldots, X_n, Y_n])$$
where $C'$ has arity 4.
By the previous propositions, $\CONTIG$ and $\leq_{lex}$ are contractible.

Similarly, 
we can define a weak version of $\GCC$ where there are no lower bounds
$\GCC(\vec{v}, \vec{0}, \vec{u}, [X_1, \ldots, X_n])$ as
$\bigwedge_{v_i \in \vec{v}} \SPLASH(C'_i, \vec{X})$,
where $C'_i$ has arity $u_i + 1$ and states that not all its arguments are equal to $v_i$.
By the previous propositions, this weak form of $\GCC$ is contractible.

However, it is notable that the $\REGULAR$ constraint is not contractible,
despite the implementation in terms of $\SLIDE$ outlined in \cite{slide}.

\begin{example}
Let $\cA$ be an automaton that accepts the language $a+b^2$.
Then $\REGULAR(\cA, [X_1]) \rightarrow X_1 = a$ but
$\REGULAR(\cA, [X_1, Y]) \rightarrow X_1 = b$.
Thus $\REGULAR$ is not contractible.
\end{example}

The discrepancy arises because $\REGULAR$ is not constructed from the operations in
the above propositions.
Essentially, the implementation defines
\[
\begin{array}{l}
\REGULAR(\cA, [X_1, \ldots, X_n]) \leftrightarrow \\
\hspace{0.9cm}
\exists \vec{Q} ~ 
\SLIDE_2(Transition, [Q_0, X_1, Q_1, \ldots, X_n, Q_n]) \\
\hspace{1.5cm}
\wedge ~ Start(Q_0) \wedge Final(Q_n)
\end{array}
\]
where the 3-ary constraint $Transition$ expresses the state transitions of $\cA$,
$Start$ defines the start state(s) and $Final$ defines the final state(s).
It is the constraint on the final variable $Q_n$ that leads to uncontractibility;
the remainder is expressible within the algebra.

We now make a simple observation that provides
a useful characterization of contractible constraints.
If ${\cal A}$ defines a prefix-closed language then $\REGULAR({\cal A}, \vec{X})$ is contractible.
This claim holds more generally.

\begin{proposition}  \label{prop:prefixclosed}
Let $C(\vec{X})$ be a constraint over a sequence of variables.
Then $C$ is contractible iff $L_C$ is prefix-closed.
\end{proposition}
\skipit{
\begin{proof}
Suppose $C$ is contractible.
If $\sigma$ is a solution of $C([X_1, \ldots, X_n, Y])$ then,
by contractibility,
the restriction of $\sigma$ to $X_1 \ldots X_n$ is a solution of $C([X_1, \ldots, X_n])$.
Thus the set of solutions is prefix-closed.

Suppose $S$ is prefix-closed.
For any solution $\sigma$ of $C([X_1, \ldots, X_n, Y])$
we know that 
the restriction of $\sigma$ to $X_1 \ldots X_n$ is a solution of $C([X_1, \ldots, X_n])$.
Since this holds for any solution $\sigma$, we have
$C([X_1, \ldots, X_n, Y]) \rightarrow  C([X_1, \ldots, X_n])$,
that is, $C$ is contractible.
\end{proof}
}

This result applies to constraints based on formal languages,
such as $\REGULAR$ and $\CFG$,
but it also applies to constraints that are formulated differently.
Thus,
for example, the solutions of $\SEQUENCE$ and $\ALLDIFF$ are prefix-closed.
Conversely, we see that constraining the final variable in a sequence, as in $Final(Q_n)$,
is not contractible.

This characterization allows us to substantiate the claim, made earlier, that
in general the negation or
implication of contractible constraints is not contractible.

\begin{example}   \label{ex:negimp}
Suppose we have an alphabet $\{a, b\}$.
If $L_C$ is $a^*$ then $L_{\neg C}$ contains $aab$, but not its prefix $aa$.
Hence $\neg C$ is not contractible.
Hence, also, ${C \rightarrow false}$ is not contractible
that is, implication of contractible constraints is not, in general, contractible.
To take another example,
if $L_{C_1}$ is $a^*b^*a^*$ and $L_{C_2}$ is $a^*b^*$ then
$L_{C_1\rightarrow C_2}$ contains $bab$ (since $bab \notin L_{C_1}$), but not its prefix $ba$
(since $ba \in L_{C_1}$ but $ba \notin L_{C_2}$).
Hence ${C_1\rightarrow C_2}$ is not contractible.
\end{example}

We can use the prefix-closed characterization both to determine whether a constraint is
contractible or not, and as the basis for approximations of uncontractible constraints.
We explore these possibilities in the following sections.

\section{Classifying Constraints}   \label{sect:class}

It is not within the scope of this paper to determine the contractibility of every global constraint.
Nevertheless, we can outline and demonstrate some principles that
make it easy, in most cases,
to classify a global constraint as contractible or not.

In general,
constraints based on counting with a lower bound (or equality) are not contractible.
We can see this by noting that any non-trivial lower bound on
the number of things in a sequence (or satisfied by a sequence)
may be violated by a prefix of the sequence.
This was already touched upon in \cite{dynamic},
where the $\SUM$ constraint $\sum_{i=1}^n X_i = N$
was shown to be non-monotonic,
but the argument holds for a wide range of constraints.

For example, $\PEAK$ counts the number of peaks in a sequence,
but a prefix of the sequence may have fewer peaks.
Similarly, $\STRETCH$ places lower bounds on the span of stretches,
so that $1 1 2 2$ might be a solution, while $1 1 2$ is not.
By a similar argument, constraints identifying properties of an extreme element in a sequence,
such as $\HIGHESTPEAK$,
are not contractible. 
On the other hand $\NOPEAK$ is contractible since,
to the extent that there is counting, there is no lower bound -- only an upper bound of 0.

We can generalize and formalize these observations.
A function $f$ is a \emph{non-decreasing accumulation function}  
if it maps sequences of values to numbers
such that, for every sequence $\vec{X}$ and value $Y$,
$f(\vec{X}Y) \geq f(\vec{X})$.
We can similarly define the non-increasing functions.
Among non-decreasing accumulation functions are 
counting the number of elements in a sequence with a fixed property, 
counting the number of different elements,
identifying the highest peaks, and
summing (some) non-negative elements of a sequence.
Note that summing possibly negative elements of a sequence is not non-decreasing.
The first part of the following proposition is an almost direct consequence of
the definitions of contractibility and non-decreasing function.
 
\begin{proposition}  \label{prop:class}
Let $C$ be a global constraint.
\begin{itemize}
\item
Suppose $C$ can be expressed as $f(\vec{X}) \leq Z$.
Then $C$ is contractible iff $f$ is a non-decreasing accumulation function.
\item
Suppose $C$ can be expressed as $f(\vec{X}) \geq Z$.
Then $C$ is contractible iff $f$ is a non-increasing accumulation function.
\item
Suppose $C$ can be expressed as $f(\vec{X}) = Z$.
Then $C$ is contractible iff $f$ is a constant function.
\end{itemize}
\end{proposition}
\skipit{
\begin{proof}
If $f$ is a non-decreasing accumulation function,
whenever $f(\vec{X}Y) \leq Z$ we must have $f(\vec{X}) \leq Z$.
Thus $C(\vec{X}Y) \rightarrow C(\vec{X})$.

If $f$ is a not a non-decreasing accumulation function,
there is a sequence of values $\vec{X}$ and a value $Y$ such that
$f(\vec{X}Y) < f(\vec{X})$.
Choose $Z$ such that $f(\vec{X}Y) \geq Z > f(\vec{X})$.
Then $C(\vec{X}Y)$ holds but $C(\vec{X})$ does not.
Thus $C$ is not contractible.

The proof of the second and third parts is similar.
\end{proof}
}

Thus the constraints $\sum_{i=1}^n X_i = N$ and $\sum_{i=1}^n |X_i| = N$ are not contractible.
Similarly, $\sum_{i=1}^n X_i \geq N$ is not contractible while
$\sum_{i=1}^n |X_i| \leq N$ is contractible.
This result can be used to establish that $\PEAK$, and $\HIGHESTPEAK$
are not contractible and that $\NOPEAK$ is contractible,
but it also applies to many other counting and summing constraints in \cite{GCcatalog}.

Notice that in constraints like $\SEQUENCE$ and $\SSUM$
the use of a lower bound in the description of the constraint $C'$ to which $\SLIDE$
is applied does not prevent contractibility.  
Each lower bound applies only to a small part of the sequence.
However, the $\RSSUM$ constraint, 
which weakens the $\SSUM$ constraint by putting bounds on the number of times
the $C'$ constraint is satisfied, is not contractible,
because 
counting is an accumulation function that is not non-increasing
and the lower bound applies to the entire sequence.

Some constraints can be recognised as contractible, based only on their informal semantics.
For example,
$\DIFFN$ and $\DISJUNCTIVE$ enforce that objects represented by the variables
are non-overlapping.
Clearly, if $\vec{X}Y$ forms a non-overlapping set, then so does $\vec{X}$ alone.
Thus contractibility follows directly from Definition \ref{defn:cont}.
Similarly,
$\CUMULATIVE$\footnote{
Under the assumption that activities can only consume resources (and not produce resources).
},  $\BINP$ and $\DISJOINT$ are contractible.

For other constraints, their informal semantics lead easily to counterexamples
to contractibility.
Constraints that involve computing the
minimum, maximum, 
mean/average, median, mode, standard deviation, etc
of the sequence
are not contractible.
This is easily recognised
since these statistics are not, in general, preserved after eliminating part of the sample set,
and hence are not prefix-closed.
Alternatively, we could recognise that these functions are not non-increasing, nor non-decreasing
and apply Proposition \ref{prop:class}.

The idea of contractibility is not useful for all global constraints.
For example,
it appears irrelevant to cyclic constraints like
the cyclic $\REGULAR$, cyclic $\SEQUENCE$ and cyclic $\STRETCH$ constraints.
In these constraints
the sequence of variables is representing a cycle or circular list and there is no natural end
at which to add variables.
Thus it is not surprising that these constraints are not contractible.

There is sometimes a fine line between contractible and uncontractible constraints.
For example,
while $\leq_{lex}$ is contractible, $<_{lex}$ is not.
To see the latter, observe that $1 1 1 <_{lex} 1 1 2$,
but the corresponding prefixes are not strictly smaller -- they are equal.
If the precedence constraint $s \prec_{\vec{X}} t$ also required that $t$ appear in $\vec{X}$,
then the constraint would not be contractible
(because $r s t$ satisfies this constraint, but $r s$ does not).
Finally, notice that the $\SEQ$ constraint is contractible, but it has the form $\SLIDE(C', \vec{X})$
where $C'$ is essentially a fixed-arity $\AMONG$ constraint;
however, the (variable-arity) $\AMONG$ constraint is \emph{not} contractible.

A quick survey of \cite{GCcatalog} suggests that most current global constraints
are not contractible,
although we have noted several useful constraints that are contractible.
In the next section we address how to propagate uncontractible open constraints.

\section{Approximating Constraints}  \label{sect:approx}

When a constraint is not contractible, the closed propagator for that constraint
is unsound as a propagator for the open constraint.
However,
following a proposal of \cite{dynamic},
we can implement an uncontractible open constraint $C(\vec{X})$ by executing 
a safe contractible approximation $C_{app}$ of $C$ until $\vec{X}$ is closed,
and then replacing $C_{app}$ by $C$ for the remainder of the execution.
To employ this approach we need to identify a contractible language containing
the language of $C$, and a propagator $C_{app}$ that implements it.

A language $L$ is an approximation of a constraint $C$ if $L_C \subseteq L$.
An approximation $L$ is contractible iff $L$ is prefix-closed.
A contractible approximation $L_a$ to a language $L$ is \emph{tight} if
for all contractible languages $L'$, if $L_a \supseteq L' \supseteq L$  then $L' = L_a$.
By Proposition \ref{prop:prefixclosed},
there is a unique contractible approximation that is tighter than all others:
the prefix-closure of $L_C$ gives the tightest contractible approximation.\footnote{
Consequently, tight and tightest contractible approximations are synonyms.
}

The prefix-closure $P(L)$ of a language $L$ often appears to be simpler than $L$.
For example, if $L_1$ is $\{a^{n^2}~|~n\in\NN\}$ then $P(L_1)$ is $a^*$.
But in general the prefix-closure is no simpler than the original language.
For example, if $L_2$ is $\{a^{n^2}b~|~n\in\NN\}$ then $P(L_2)$ is $a^* \cup L_2$.
In some cases it is easy to represent $P(L)$ when given a representation of $L$.
In particular,
when $L$ is defined by a finite automaton
the automaton accepting $P(L)$ is easily computed.

\begin{proposition}   \label{prop:prefixFA}
Let $\cA$ be a (possibly nondeterministic) finite state automaton, and
let $\cA'$ be the finite state automaton obtained from $\cA$ by
making final all states on a path from the start state to a final state.
Then $L(\cA') = P(L(\cA))$.
$\cA'$ can be computed in linear time.
\end{proposition}
\skipit{
\begin{proof}
Consider any prefix $w$ of a word $wu \in L(\cA)$.
$wu$ describes a path in $\cA$ that ends at a final state.
Hence $w$ describes a path in $\cA$ that ends at a state on a path to a final state.
Hence $w$ is accepted by $\cA'$.
Thus $L(\cA') \supseteq P(L(\cA))$.

Conversely, suppose $w$ is accepted by $\cA'$.
By the construction of $\cA'$, $w$ describes a path in $\cA$ that ends at a state $Q$
on a path to a final state of $\cA$.
Let $u$ be a word corresponding to a path from $Q$ to a final state.
Then $wu$ is accepted by $\cA$ and hence $w$ is a prefix of a word in $L(\cA)$.
Thus $L(\cA') \subseteq P(L(\cA))$.

We can construct $\cA'$ as follows.
Treat the automaton $\cA$ as a directed graph with the states as vertices and where
each transition from $Q_1$ to $Q_2$ is represented by an edge from $Q_1$ to $Q_2$.
Perform depth-first search and mark all states reachable from the start state.
Now consider the graph with the edges reversed.
Perform depth-first search from the reachable final states, 
marking each visited reachable state as a final state.
$\cA'$ is the automata $\cA$ with these additional final states.
The cost of the construction is O($V+E$),
where $V$ is the number of states and $E$ is the number of transitions.
(Note that we could ignore reachability and define a variation of $\cA'$
that may have some unreachable final states.)
\end{proof}
}

Similarly, we can use the structure of a context-free grammar to construct
a grammar for its prefix-closure.

\begin{proposition}   \label{prop:prefixCFG}
Given a context-free grammar $\cG$ defining a language $L$,
a context-free grammar $\cG'$ for $P(L)$ can be generated in quadratic time,
and in linear time if $\cG$ is in Chomsky normal form.
\end{proposition}
\skipit{
\begin{proof}
(Sketch)
We show only the construction when $\cG$ is presented in Chomsky normal form,
and leave the generalization to arbitrary grammars and
the verification of its correctness to the reader.

Let $\cG = (N, T, R, S)$,
where $N$ is a set of nonterminal symbols,
$T$ is a set of terminal symbols,
$R$ is the set of production rules, and
$S$ is the start symbol.
In Chomsky normal form, production rules have the form
$A \rightarrow BC$ or $A \rightarrow a$ or $S \rightarrow \varepsilon$
where $A$, $B$, and $C$ are nonterminal symbols, $a$ is a terminal symbol,
and $\varepsilon$ is the empty word.
We define
$\cG' = (N', T, R', S')$, where
$N' = N \cup \{ S' \} \cup \{ A_p ~|~ A \in N \}$
and
\[
\begin{array}{rl}
R' = R ~ \cup & \{ S' \rightarrow \varepsilon \} \cup \{ S' \rightarrow S_p \}
\cup 
 \{ A_p \rightarrow a ~|~ (A \rightarrow a) \in R \}  ~ \cup  \\
&  \{ A_p \rightarrow B_p ~|~ (A \rightarrow BC) \in R \} \cup 
 \{ A_p \rightarrow BC_p ~|~ (A \rightarrow BC) \in R \} 
\end{array}
\]
For each nonterminal $A \in N$,
$A_p$ generates all non-empty prefixes of words generated by $A$,
including the words generated by $A$.
It is clear that $\cG'$ is larger than $\cG$ by a factor of 3 or less.
For an arbitrarily structured grammar, the size of $\cG'$ can grow quadratically.


$R'$ is not in Chomsky normal form, but it is easily simplified to that form.
Nonterminals $A_p$ which are strongly connected by edges corresponding to productions of the form
$X \rightarrow Y$ can be replaced by a single equivalent nonterminal, to give $R''$.
Remaining productions $X \rightarrow Y$ can be replaced by a set of productions
$\{ X \rightarrow \psi ~|~ (Y \rightarrow \psi) \in R'' \}$.
In general, repeated replacements are necessary to eliminate all $X \rightarrow Y$  productions.
A naive representation can increase the size of the grammar,
but a more careful representation can share the right-hand side of productions
so that the Chomsky normal form is not larger than $\cG'$.
\end{proof}
}

Thus, the tightest contractible approximation of 
$\REGULAR(\cA, \vec{X}, N)$ is implemented by
$\REGULAR(\cA', \vec{X}, N)$, and
the tightest contractible approximation of 
$\CFG(\cG, \vec{X}, N)$ is implemented by
$\CFG(\cG', \vec{X}, N)$.

As a corollary to Proposition \ref{prop:prefixFA}, we can check in linear time whether 
a language defined by a deterministic finite automaton is prefix-closed:
we simply check whether the construction of $\cA'$ in Proposition \ref{prop:prefixFA}
made any new final states.
This improves on a result of \cite{BSX}.
Unfortunately,
recognising when a language defined by a nondeterministic finite automaton $\cA$ is prefix-closed
is not so simple;
$\cA$ need not have the property that all states on a path from start to final state are final.
It is shown in \cite{BSX} that this problem is PSPACE-complete.
The problem is undecidable for languages defined by context-free grammars \cite{BSX}.
However, the decision problem is much less important than the ability to construct 
(the representation of) the prefix-closure,
so these negative results are not significant.

$\REGULAR$ and $\CFG$ are complicated by flexible parameters,
but approximations to simpler constraints are often correspondingly simpler
to recognise.
As discussed in \cite{dynamic},
a constraint $\sum_{i=1}^n X_i = N$
where the $X_i$'s must be non-negative
is not monotonic but is approximated by the constraint 
$\sum_{i=1}^n X_i \leq N$.
Using Proposition \ref{prop:class} we can recognise this as the tightest contractible approximation.
Similarly, for a counting constraint such as $\PEAK(\vec{X}, N)$,
which states that there are exactly $N$ peaks in $\vec{X}$,
the tightest contractible approximation states that $N$ is an upper bound on the number of peaks.
In the same way, $\NVALUE(\vec{X}, N)$ is best approximated by treating $N$ only as an upper bound.
The tightest approximation of the $\GCC$ is the weak form of $\GCC$ discussed in Section \ref{sect:cont}.
In all these cases, since counting is a non-decreasing accumulation function,
the tightest contractible approximation is to eliminate the lower bounds.
In $\HIGHESTPEAK(\vec{X}, Z)$, the height of the highest peak is 
a non-decreasing accumulation function and so
the tightest approximation states that $Z$ is an upper bound on the height of the highest peak.

On the other hand,
for some constraints where the accumulation function is neither non-increasing nor non-decreasing
there appear to be no non-trivial approximations.
For example,
consider a constraint $\AVERAGE(\vec{X}, M)$ stating that $M$ is the mean/average of
the values of $\vec{X}$.
Given a fixed $M$,
\emph{any} sequence of values can be a prefix of a sequence with mean $M$.
Hence the tightest contractible approximation of $\AVERAGE$ is the constraint that accepts any sequence,
that is, the constraint $true$.
For such a constraint there is no propagation until the constraint is closed.

However, as the previous discussion shows,
for many constraints the tightest contractible approximation is not only non-trivial,
it has a clear and simple expression.
For these constraints a propagator for the approximation $C_{app}$ is almost ready-made,
given a propagator for the original constraint $C$.
Furthermore, the transition of propagator from $C_{app}$ to $C$ when the constraint closes
can be smooth and simple because, in the cases above,
the propagator for $C_{app}$ is simply a weakened form of the propagator for $C$.
Some more detailed analysis of this similarity of propagators for $C$ and $C_{app}$, for several constraints $C$,
appears in \cite{open1} and (for a slightly different model of open constraint) \cite{open2}.

If we have domain consistent closed propagators
and a tight contractible approximation, then we can obtain an open D-consistent propagator
from Bart\'{a}k's proposal.
Recall that under Bart\'{a}k's proposal \cite{dynamic},
a closed propagator for $C_{app}$ is dynamised to handle extensions of the sequence of variables
(possibly through his generic dynamisation).
This propagator is then executed until the sequence of variables is closed,
at which point the propagator is replaced by a closed propagator for $C$.

\begin{theorem}  \label{thm:openDcons}
Let $C_{app}$ be the tightest contractible approximation to $C$,
and suppose we have closed propagators for $C_{app}$ and $C$
that maintain domain consistency for $\vec{X}$.
Then Bart\'{a}k's proposal maintains
open D-consistency for $C$.
\end{theorem}
\skipit{
\begin{proof}
Since $C_{app}$ is contractible,
domain consistency of $C_{app}$ for $\vec{X}$ is equivalent to open D-consistency on $C(\vec{X})$.
This follows because $C_{app}$ is the prefix-closure of $C$ and so
every support for domain consistency of $C_{app}(\vec{X})$ for $\vec{X}$
corresponds to a longer word that is a support for D-consistency on $C(\vec{X})$,
and \emph{vice versa}
every support for open D-consistency on $C(\vec{X})$ has a corresponding prefix
that is a support for domain consistency of $C_{app}(\vec{X})$ for $\vec{X}$.
Once $\vec{X}$ is closed, 
domain consistency for $\vec{X}$ is identical to D-consistency on $C(\vec{X})$.
\end{proof}
}

We can obtain similar results for consistency conditions other than domain consistency.
All that is required is to define the appropriate corresponding open consistency.
For example, consider bounds consistency.
Let $min(X)$ ($max(X)$) denote the smallest (largest) value in $D(X)$.
The appropriate form of bounds consistency for open constraints is open B-consistency.
\begin{definition}
Given a domain $D$,
an occurrence of a constraint $C(\vec{X})$ is
\emph{open B-consistent} if

for every $X_i \in \vec{X}$, 
and for $d_i = min(X_i)$ and $d_i = max(X_i)$,
there is a word $d_1 \ldots d_m$ in $L_C$ such that
$|\vec{X}| \leq m$,
and
$d_j \in min(X_i)..max(X_i)$ for $j = 1,\ldots,|\vec{X}|$.
\end{definition}

We can now express the corresponding result for bounds consistency.
The proof is essentially the same as that for the previous theorem.

\begin{corollary}  \label{cor:openBcons}
Let $C_{app}$ be the tightest contractible approximation to $C$,
and suppose we have closed propagators for $C_{app}$ and $C$
that maintain bounds consistency for $\vec{X}$.
Then Bart\'{a}k's proposal maintains
open B-consistency for $C$.
\end{corollary}

Notice that we still require a tightest contractible approximation.
Any weakening of this requirement can lose open B-consistency,
as is clear from Corollary \ref{cor:cont}.

\section{Contractibility of Soft Constraints}    \label{sect:softclass}

We consider ``soft'' global constraints in the style of \cite{softcons}\nofootnote{
Consideration of other forms of soft constraint are left for future research.}.
In such constraints there is a violation measure\footnote{
Also called violation cost \cite{softcons}.
},
which measures the degree to which an assignment to the variables
violates the associated ``hard'' constraint,
and solutions are assignments that satisfy an upper bound on the violation measure.
Thus such soft constraints have the form $m(\vec{X}) \leq Z$,
where $m$ is the violation measure.
We refer to the hard constraint as $C(\vec{X})$ and 
the corresponding soft constraint as $C_s(\vec{X}, Z)$.

Assessing the contractability of such constraints is made easier by Proposition \ref{prop:class},
which says that a constraint $m(\vec{X}) \leq Z$ is contractible iff $m$ is non-decreasing.
Given this characterization,
we will refer to non-decreasing accumulation functions as contractible functions.
To evaluate whether or not soft constraints are contractible
we must consider the form of the violation measure,
and whether it forms a contractible function.

\begin{definition}
A \emph{violation measure} for a sublanguage $L$ of a language $L'$
is a function $m$ which maps $L'$ to 
the non-negative real numbers, such that if $w \in L$ then $m(w) = 0$.
$m$ is \emph{proper} for $L$ if
for all words $w \in L'$, $m(w) = 0$ iff $w \in L$.
A violation measure for a constraint $C(\vec{X})$ is a violation measure for $L_C$
as a sublanguage of the static type $T(\vec{X})$.
\end{definition}

For example,
a use of $\ALLDIFF$ might give the set $\ZZ$ of integers as the static type of each variable.
A violation measure might then be
the number of disequalities $X_i \neq X_j, i \neq j$ violated by a valuation for $\vec{X}$,
or the number of variables equal to another variable under the valuation,
or  the minimum absolute value of the sum over $i$ of values $c_i$ such that,
for each $i$ and $j$ with $i \neq j$, $X_i + c_i \neq X_j + c_j$.\footnote{
This latter measure expresses the smallest perturbation $\vec{c}$ of the values for the variables needed to satisfy
the $\ALLDIFF$ constraint.
More formally, 
$m(\vec{X}) = \min_{\vec{c}} \{ |\sum_{i=1}^n c_i | ~|~ \forall j ~ j \neq i \rightarrow X_i + c_i \neq X_j + c_j \}$.
}
It is easy to see that each of these defines a violation measure.
The third is not a proper violation measure because, for example,
the word $11233$ can have perturbations $c_i$ of $0,-1,0,0,1$.
Thus $m(11233) = 0$ but $11233 \not\in L_C$.
(Summing the absolute value of the $c_i$, on the other hand, would lead to a proper measure.)

Proper violation measures for a language $L$
are a refinement of the characteristic function of $L$.\footnote{
Indeed, for any proper violation measure $m$,
the corresponding hard constraint can be recovered as $m(\vec{X}) \leq 0$.
}
Most violation measures in the literature are proper for their intended language.
Although any function from words to non-negative reals can be considered a proper violation measure
by appropriate choice of language $L$,
in practice the hard constraint determines $L$ and 
the violation measure is then designed to be proper.
A non-proper measure can be considered misleading because a word $w$
that violates the language $L$ can have a violation measure of 0.
We admit non-proper violation measures mainly because
contractible approximations considered in Section \ref{sect:softapprox} can be non-proper.
However, we make some effort in this section to identify proper violation measures.

There are three broad classes of violation measures \cite{open3}:
those based on constraint decomposition, edit distance, and graph properties.
We address the first two classes in the following subsections.
The richness of the graph property framework  \cite{costeval} makes it difficult to obtain
broad results on contractibility.
A somewhat narrow sufficient condition for contractability of soft constraints
defined by graph property-based violation measures is presented in \cite{open3}.
For each of the classes we consider,
we will incorporate a weighting that adds greater flexibility and expressiveness
to the class.

\subsection{Decomposition-based Violation Measures}   \label{sect:decomp}

Many hard constraints can be decomposed into elementary constraints,
whether naturally (such as the decomposition of $\ALLDIFF$ into disequalities)
or by a construction, as in \cite{decomp}.
Violation measures can be constructed by combining the violations of each elementary constraint.
We define a general class of decomposition-based violation measures  that  includes as
special cases:
primal graph based violation costs \cite{softcons},
decomposition-based violation measures of \cite{warming},
the value-based violation measure for $\GCC$ \cite{softcons,warming},
the measures used for the soft $\SEQUENCE$ constraint \cite{MNQW}
and the soft $\CUMULATIVE$ constraint \cite{softcum},
the weighted measures for $\ALLDIFF$ and $\GCC$ \cite{MBL07,MBL09},
and the class of decomposition-based measures discussed in \cite{open3}.
We begin with several definitions.

A \emph{weighted set} is a pair $(S, w)$ where $S$ is a set and
$w$ is a function mapping each element of $S$ to a non-negative real number or $\infty$.
Values not in $S$ have weight $0$.
If these are the only values of weight $0$ we say $(S, w)$ is \emph{proper}.
A weighted set is a minor generalization of a multiset.
A weighted set $(S_1, w_1)$ is a \emph{sub-weighted set} of weighted set  $(S_2, w_2)$ if,
for every element $s \in S_1$, $w_1(s) \leq w_2(s)$.
Union of weighted sets is defined by
$
(S_1, w_1) \cup (S_2, w_2)  = (S_1 \cup S_2, w_1 + w_2)
$
where $(w_1+w_2)(x) = w_1(x) + w_2(x)$.
When a weighted set contains things with variables that are subject to substitution,
the application of a substitution might unify elements of the set.
Hence, $(S, w)\theta$ denotes $(S\theta, w')$ where
$w'(s)$ is the sum of $w_1(s')$ over all $s' \in S$ such that $s'\theta \equiv s$.

We need to carefully formalize the notion of decomposition.
The definition takes as a parameter a class of elementary constraints.
Usually the constraints in such a class have bounded arity.
\begin{definition}   \label{defn:decomp}
A \emph{decomposition} is a function that maps a constraint $C$ with a given type $T$
and a sequence of variables $\vec{X}$
to a tuple $(\vec{X}, \vec{U}, T', S, w)$
where 
$\vec{U}$ is a collection of new variables,
$T'$ is an extension of $T$ to $\vec{U}$,
and
$(S, w)$ is a proper weighted set of elementary constraints over $\vec{X} \vec{U}$
such that
$C(\vec{X}) \leftrightarrow \exists \vec{U} ~ T'(\vec{U}) \wedge \bigwedge_{s \in S} s$.
\end{definition}
The weights in this definition are used only to emphasize some constraints in a decomposition over others;
in particular, the infinite weight allows us to specify elementary constraints that must not be violated.
An \emph{unweighted decomposition} is one where all constraints in $S$ have 
the same, non-zero weight.
In that case, we may omit $w$.
We write $\decomp(C(\vec{X}))$ to express the weighted set $(S, w)$, or simply $S$ when the decomposition is unweighted.

This definition of decomposition is very broad, perhaps too broad,
since it allows the set of elementary constraints and/or their weights to vary radically 
as the length of $\vec{X}$ changes.
For example, it permits using the decomposition of $\ALLDIFF(\vec{X})$ into disequalities
when $|\vec{X}|$ is odd, and a decomposition from \cite{decomp} (see Example \ref{ex:rc_alldiff})
when $|\vec{X}|$ is even.
However, we will see in Example \ref{ex:rs} a constraint whose expression requires 
some of the flexibility offered by this broad definition.

An \emph{error function} $e$ maps an elementary constraint and a valuation to a non-negative real number,
representing the amount of error (or violation) of the constraint by the valuation.
We require that $e(v, c) = 0$ iff $c$ is satisfied by $v$.
We extend $e$ to weighted sets of constraints by defining $e(v, (S,w)) = (S', w')$
where $S' = \{e(v, s) ~|~ s \in S\}$ and $w'(x) = \sum_{s \mid v(s)=x} w(s)$.

A \emph{combining function} maps a weighted set of numbers to a single number.
A combining function $comb$ is \emph{monotonic} if,
whenever $(S_1, w_1)$ is a sub-weighted set of $(S_2, w_2)$, $comb (S_1, w_1) \leq comb (S_2, w_2) $.
The function $comb$ is \emph{disjunctive} if for all weighted sets of reals $(S, w)$,
$comb(S, w) = 0$ iff $S = \{ 0 \}$.
We say $comb$ \emph{has unit 0} if,
for every $(S, w)$ and $w'$, $comb((S, w) \cup (\{0\}, w')) = comb(S, w)$.
Counting non-zero values, summation, sum of squares, and maximization are examples of monotonic,
disjunctive combining functions with unit 0;
product and minimization are neither monotonic nor disjunctive nor have unit 0.

\begin{definition}  \label{defn:dbvm}
A \emph{decomposition-based violation measure} $m$ for a constraint $C(\vec{X})$
with type $T$
is based on a decomposition  $(\vec{X}, \vec{U}, T', S, w)$ of $C(\vec{X})$,
an error function $e$,
and
a combining function $comb$
and is defined by, for each valuation $v$ of $\vec{X}$,

\[
m(v(\vec{X})) = \min_{v'} comb( e(v', \decomp(C(\vec{X}))))
\]
where we minimize over all extensions $v'$ of $v$ to $\vec{U}$ that satisfy $T'$.
\end{definition}

This definition was inspired by
the formulation of hierarchical constraints in \cite{CH,HCLP}.
The violation counting decomposition measures of \cite{softcons,warming}
can be obtained 
when the error function $e(v, c)$ returns 0 if $v$ satisfies $c$ and 1 otherwise,
and the combining function is summation.
The value-based measures of \cite{softcons,warming,MNQW,softcum} also use summation as the combining function,
but use an error function that returns the amount by which the constraint $c$ is violated by
the valuation $v$.
If we use maximization or the sum of squares in place of summation we have new violation measures
similar to the \emph{worst-case-better} and \emph{least-squares-better} comparators of \cite{CH,HCLP}.
Clearly many violation measures are available for a constraint by making different choices
for the decomposition and the error and combining functions.

There is a powerful sufficient condition for a decomposition-based violation measure to be proper. 

\begin{proposition}
Let $m$ be a decomposition-based violation measure for a constraint $C$,
as defined in Definition \ref{defn:dbvm} with combining function $comb$.
$m$ is proper for $L_C$ if
$comb$ is disjunctive.
\end{proposition}
\skipit{
\begin{proof}
Let $v$ be a valuation for $\vec{X}$.
Suppose $comb$ is disjunctive.

\noindent
$m(v(\vec{X})) = 0$

\noindent
iff
$\min_{v'} comb( e(v', \decomp(C(\vec{X})))) = 0$

\noindent
iff for some $v'$ extending $v$,
$comb( e(v', \decomp(C(\vec{X})))) = 0$

\noindent
iff for some $v'$ extending $v$,
and some $w$,
$e(v', \decomp(C(\vec{X}))) = (\{0\}, w)$

\noindent
iff for some $v'$ extending $v$,
$v'$ satisfies every $c \in \decomp(C(\vec{X}))$

\noindent
iff
$v$ satisfies $C(\vec{X})$

\noindent
iff
$v(\vec{X}) \in L_C$.

Thus, for any valuation $v$,
$m(v(\vec{X})) = 0$ iff $v(\vec{X}) \in L_C$.
Hence, $m$ is proper for $L_C$.
\end{proof}
}

We now turn to the problem of recognizing contractibility.
We say that one formula $(\vec{X}, \vec{U}, T_1, S_1, w_1)$ is \emph{covered} 
by another formula $(\vec{W}, \vec{V}, T_2, S_2, w_2)$ 
if 
there is a substitution $\theta$ that maps 
$\vec{X}$ into $\vec{W}$ and
$\vec{U}$ into $\vec{V} \cup \vec{W} \cup \Sigma$,
where $\Sigma$ is a set of constants,
such that 
$T_1(\vec{X}) = T_2(\vec{X}\theta)$,
$(S_1, w_1) \theta$ is a sub-weighted set of  $(S_2, w_2)$\nofootnote{
A weighted set $(S_1, w_1)$ is a \emph{sub-weighted set} of weighted set  $(S_2, w_2)$ if,
for every element $s \in S_1$, $w_1(s) \leq w_2(s)$.
$(S_1, w_1)\theta$ denotes $(S_1\theta, w')$ where
$w'(s)$ is the sum of $w_1(s')$ over all $s' \in S_1$ such that $s'\theta \equiv s$.
}
and $T_2(\vec{U}\theta) \subseteq T_1(\vec{U})$.
Covering has some similarity to characterizations of
containment of conjunctive relational database queries \cite{CM77}, 
(constraint) logic programming rule subsumption \cite{equiv,lift}, 
and sufficient conditions for query containment under bag semantics \cite{ChauVardi,IoannRama}.

\begin{example}  \label{ex:decomp1}
The decomposition of $\ALLDIFF(\vec{X})$ into an unweighted set of disequalities is formalized as
$(\vec{X}, \emptyset, T, S, w)$
where $S$ is the set of disequalities and $w$ gives every disequality a weight of $1$.
It is clear that the decomposition of $\ALLDIFF(\vec{X})$ is covered by that of $\ALLDIFF(\vec{X}Y)$
where the substitution is the identity.
\end{example}

\begin{example}  \label{ex:decomp2}
$\CONTIG$ is implemented in \cite{contig} essentially by the decomposition
\[
\CONTIG(\vec{X}) \leftrightarrow
\exists \vec{L}, \vec{R}
 ~  \bigwedge_{i=2}^{n-1} C'( X_{i-1}, R_{i-1}, L_i, X_i, R_i, L_{i+1}, X_{i+1})
 \]
for a constraint $C'$.
This decomposition is formalized as
$(\vec{X}, \vec{L}\vec{R}, T, S, w)$
where $T$ gives all variables a type of $\{0, 1\}$,
$S$ is the set of $C'$ constraints,
and $w$ gives every constraint a weight of $1$.
Alternatively, if contiguity is more important for variables nearer the right end of the sequence $\vec{X}$,
we might weight each $C'$ constraint by the largest index of a variable appearing in it.
The decomposition of $\CONTIG(\vec{X}Y)$ covers that of $\CONTIG(\vec{X})$
where the substitution is the identity on $\vec{X}$, $\vec{L}$, and $\vec{R}$.
\end{example}

We can now provide a sufficient condition for 
a soft constraint with a decomposition-based violation measure
to be contractible.

\begin{proposition}  \label{prop:softdecomp}
Let $C_s$ be a soft constraint with a decomposition-based violation measure
defined using a monotonic combining function.
Let $(\vec{X}, \vec{U}, T_1, S_1, w_1)$ be the decomposition of $C(\vec{X})$
and $(\vec{X}Y, \vec{V}, T_2, S_2, w_2)$ be the decomposition of $C(\vec{X}Y)$.
If $(\vec{X}, \vec{U}, T_1, S_1, w_1)$ is covered by $(\vec{X}Y, \vec{V}, T_2, S_2, w_2)$ 
via a substitution that is the identity on $\vec{X}$
then $C_s$ is contractible.
\end{proposition}
\skipit{
\begin{proof}
By the covering condition, there is a substitution $\theta$ 
that is the identity on $\vec{X}$
and maps $\vec{U}$ to $\vec{V} \cup \vec{X}Y \cup \Sigma$
such that $(S_1, w_1) \theta$ is a sub-weighted set of  $(S_2, w_2)$.
Consider any assignment $v$ to $\vec{X}Y \cup \vec{V}$.
Then $v \circ \theta$ is an assignment\footnote{
We define $(v \circ \theta)(x) = v(x\theta)$ for any term $x$.
} to $\vec{X} \cup \vec{U}$.
Furthermore, 
$v((S_1, w_1) \theta)$ is a sub-weighted set of  $v(S_2, w_2)$
and hence
$e(v \circ \theta, (S_1, w_1)) = e(v, (S_1, w_1) \theta)$ is a sub-weighted set of  $e(v, (S_2, w_2))$.
Consequently, since the combining function $comb$ is monotonic,
$comb(e(v \circ \theta, (S_1, w_1))) \leq comb(e(v, (S_2, w_2)))$.
It follows that $m(v(\vec{X})) \leq m(v(\vec{X}Y))$.
Thus, since $v$ is arbitrary, $m$ is non-decreasing and, by Proposition \ref{prop:class},
$C_s$ is contractible.
\end{proof}
}

It follows that the constraints in Examples \ref{ex:decomp1} and \ref{ex:decomp2} are contractible.
More generally, 
if an unweighted decomposition is defined via part of the algebra discussed in Section \ref{sect:cont}
(that is, using
$\SLIDE$ or $\SPLASH$ meta-constraints,
constraints on a fixed finite prefix of the variable sequence, conjunction and existential quantification)
and a monotonic combining function
then Proposition \ref{prop:softdecomp} is sufficient to establish contractibility.
However, covering is not a necessary condition for contractibility, as the following example demonstrates.

\begin{example}   \label{ex:rs}
Consider the definition of a rising sawtooth relation $rs$ on variables $\vec{X}$.
In such a relation, the subsequence of values in even numbered positions forms a non-decreasing sequence,
and every value in odd numbered positions is greater than or equal to its immediately adjacent neighbours.
\footnote{
This is an artificial constraint, designed to demonstrate the point.
However, the pricing of goods with volume discounts can have a similar rising sawtooth behaviour.
}
This relation can be decomposed into elementary constraints as follows.
The decomposition is defined recursively, but notably requires two recursive cases,
corresponding to the distinction between odd and even length sequences.
\[
\begin{array}{ll}
\decomp(rs([])) & = true \\
\decomp(rs([X_1])) & = true \\
\decomp(rs([X_1, X_2])) & = X_1 \geq  X_2 \\
\decomp(rs([X_1, \ldots, X_{2n}, X_{2n+1}])) & = \\
& \hspace{-4.0cm} \decomp(rs([X_1, \ldots, X_{2n}])) \wedge X_{2n+1} \geq X_{2n} \\
\decomp(rs([X_1, \ldots, X_{2n}, X_{2n+1}, X_{2n+2}])) & = \\
& \hspace{-4.0cm} \decomp(rs([X_1, \ldots,  X_{2n}])) \wedge X_{2n+1} \geq X_{2n+2} \wedge X_{2n+2} \geq X_{2n}
\end{array}
\]
Consider the soft constraint derived from this decomposition by counting the number of violations.
It is clear that the sufficient condition of Proposition \ref{prop:softdecomp}
does not apply because there is no covering.
Nevertheless, we can verify that a decomposition-based soft $rs$ constraint is contractible.
Note first that when $\vec{X}$ has even length
$\decomp(rs(\vec{X})) \subseteq \decomp(rs(\vec{X}Y))$
and consequently the violation measure is non-decreasing in this case.
When $\vec{X}$ has odd length the relationship is less obvious.
However, we know that
\[
(X_{2n+1} \geq X_{2n+2}) \wedge  (X_{2n+2} \geq X_{2n}) \rightarrow  (X_{2n+1} \geq X_{2n})   
\]
and its contrapositive
\[
\neg (X_{2n+1} \geq X_{2n}) \rightarrow \neg (X_{2n+1} \geq X_{2n+2}) \vee \neg (X_{2n+2} \geq X_{2n})
\]
Hence, any valuation for the variables that gives rise to a violation of $X_{2n+1} \geq X_{2n}$ 
will also give rise to a violation of  $X_{2n+1} \geq X_{2n+2}$, or  $X_{2n+2} \geq X_{2n}$, or both.
Thus the violation measure is non-decreasing in this case also.
Since the violation measure is non-decreasing, the decomposition-based soft $rs$ constraint is contractible.

Similarly, the violation measures derived from summing the amount of violation or
taking the maximum amount of violation of any elementary constraint
lead to contractible soft $rs$ constraints.
\end{example}

This example demonstrates a major limitation of the sufficient condition in Proposition \ref{prop:softdecomp}:
it addresses only the syntactic structure of the decomposition.
However some constraints, such as $rs$,
require reasoning about the semantics of the elementary constraints in order to recognise
that the decomposition-based soft constraint is contractible.
(For $rs$ we exploited the knowledge that $\geq$ forms a total order.)

{
A second example is given by a decomposition of $\ALLDIFF$ given in \cite{decomp}.

\begin{example}  \label{ex:rc_alldiff}
Consider the $\ALLDIFF$ constraint with type $T$ that maps each $X_i$ to $1..d$,
which we denote by $\ALLDIFF_T$.
To define the decomposition we need to introduce variables $A_{i l u}$ of type $\{0, 1\}$
and constraints as follows.

For $1 \leq i \leq n$ and $1 \leq l \leq u \leq d$
we have the constraints
\begin{equation}  \label{eq:diff_Ailu1}
A_{i l u} = 1 \leftrightarrow X_i \in [l, u]
\end{equation}

\begin{equation}  \label{eq:diff_Ailu2}
\sum_{i=1}^n A_{i l u} \leq u - l + 1
\end{equation}

\noindent
This decomposition is formalized as $(\vec{X}, \vec{A}, T', S, w)$
where 
$T'$ extends $T$ to the $A_{i l u}$ variables,
$S$ consists of the constraints $\eqref{eq:diff_Ailu1}$ and $\eqref{eq:diff_Ailu2}$
and $w$ gives all constraints the same weight.
It is easy to establish that
$\ALLDIFF_T(\vec{X}) \leftrightarrow \exists \vec{A}  \in T'(\vec{A}) ~ \eqref{eq:diff_Ailu1} \wedge \eqref{eq:diff_Ailu2} $.

When $\vec{X}$ is extended by $Y$, the decomposition contains extra variables
$A_{(n+1) l u}$,
extra constraints of type \eqref{eq:diff_Ailu1} involving $Y$ and the new variables,
and replaces constraints \eqref{eq:diff_Ailu2}  by

\begin{equation}   \label{eq:diff_Ailu3}
\sum_{i=1}^{n+1} A_{i l u} \leq u - l + 1
\end{equation}

Now, for each $l$ and $u$,
$\eqref{eq:diff_Ailu3} \wedge (0 \leq A_{(n+1) l u} \leq 1) \rightarrow \eqref{eq:diff_Ailu2}$.
Thus, every valuation that violates \eqref{eq:diff_Ailu2} will also violate \eqref{eq:diff_Ailu3}.
It follows that the soft constraint based on counting violations in this decomposition of $\ALLDIFF$
is contractible.
Similarly, soft constraints based on summing violation amounts or taking the maximum are also contractible,
because 
$\sum_{i=1}^{n+1} A_{i l u} \geq \sum_{i=1}^{n} A_{i l u}$.

On the other hand, 
the decomposition of
$\ALLDIFF(\vec{X}Y)$
cannot be a covering of the decomposition of
$\ALLDIFF(\vec{X})$,
because each constraint \eqref{eq:diff_Ailu2} is not covered by the corresponding constraint \eqref{eq:diff_Ailu3}.
Thus, again, the sufficient condition of Proposition \ref{prop:softdecomp} cannot be used.
\end{example}
}

To redress the weakness of covering in addressing Examples \ref{ex:rs} and \ref{ex:rc_alldiff}
we need to incorporate knowledge of the semantics of the elementary constraints and,
more generally, the error function.
We begin with some definitions.

A \emph{division} of a weighted set $(S, w)$ is a collection of sub-weighted sets $(S_i, w_i)$
such that $\cup _i ~ (S_i, w_i) = (S, w)$.
When all $S_i$ are singleton sets we refer to this as \emph{division into singletons}.
Given a weighted set $(S, w)$,
we write $w\theta$ to denote the weight function of $(S, w)\theta$.

\begin{definition}
A \emph{semantic embedding} of 
$(\vec{X}, \vec{U}, T_1, S, w)$ in $(\vec{X}Y, \vec{V}, T_2, S', w')$
is a pair $\langle \phi, \theta \rangle$, where 
$\phi$ is a function and $\theta$ is a substitution, such that
\begin{itemize}
\item
$\theta$
is the identity on $\vec{X}$ and maps
$\vec{U}$ into $\vec{X}Y \cup \vec{V} \cup \Sigma$,
where $\Sigma$ is a set of constants,
such that 
$T_2(\vec{U}\theta) \subseteq T_1(\vec{U})$;
\item
$\phi$ is an injective function from $(S, w)\theta$ to a division of $(S', w')$; and
\item
for every valuation $v$ and
every elementary constraint $c \in S\theta$,
$e(v, (\{c\}, w\theta)) \leq comb(e(v, \phi(c)))$.
\end{itemize}
\end{definition}

In a semantic embedding,
the substitution $\theta$ shows how variables local to the first decomposition are represented in the second
and the function $\phi$ shows how elementary constraints in the first decomposition are represented in the second.
The third condition requires that these representations respect the semantics expressed by the error function $e$.

Covering is essentially a syntactic form of semantic embedding:
a semantic embedding where $(S', w')$ is divided into singletons
and any constraint $c\theta$ in $S\theta$ is mapped to $c\theta$ in $S'$.

We are now in a position to state a much broader sufficient condition for contractibility
than Proposition \ref{prop:softdecomp}.

\begin{theorem}  \label{thm:decomp}
Let $C_s$ be a soft constraint with a decomposition-based violation measure $m$
defined using a monotonic combining function $comb$.
Let $(\vec{X}, \vec{U}, T_1, S_1, w_1)$ be the decomposition of $C(\vec{X})$
and $(\vec{X}Y, \vec{V}, T_2, S_2, w_2)$ be the decomposition of $C(\vec{X}Y)$.
Suppose there is a semantic embedding of $(S_1, w_1)$ in $(S_2, w_2)$.
Then $C_s$ is contractible.
\end{theorem}
\skipit{
\begin{proof}
Consider the extension of $\vec{X}$ to $\vec{X}Y$ and a valuation $v$ on $\vec{X}Y\vec{V}$.
Let $\langle \phi, \theta \rangle$ be the semantic embedding.
Then,
for every elementary constraint $c \in S_1\theta$,
$e(v, (\{c\}, w_1\theta)) \leq comb(e(v, \phi(c)))$.
Hence
$e(v \circ \theta, (S_1, w_1)) = e(v, (S_1, w_1)\theta) 
=  comb(\bigcup_{c \in S_1\theta} e(v, (\{c\}, w_1\theta))) 
\leq  comb(\bigcup_{c \in S_1\theta} e(v, \phi(c)) )
\leq  comb(e(v, (S_2, w_2)) )$.

Since $comb$ is monotonic,
$comb(e(v \circ \theta, (S_1, w_1))) \leq comb(e(v, (S_2, w_2)) )$.
It follows that
$\min_v comb(e(v, (S_1, w_1))) \leq \min_v comb(e(v, (S_2, w_2)) )$,
and hence $m( C(\vec{X}) ) \leq m( C(\vec{X}Y) )$.
Thus $C_s$ is contractible.
\end{proof}
}

For (unweighted) violation counting measures, the third condition of semantic embedding reduces to
$D \models (T_2(\vec{V}) \wedge \phi(c)) \rightarrow c\theta$,
where $D$ expresses some properties of the elementary constraints.
Thus, for these measures, we can reason about contractibility using conventional logic.
In Example \ref{ex:rs},
$\theta$ can be the identity substitution, since no additional variables are used in the decomposition,
and $\phi$ maps $(X_{2n+1} \geq X_{2n})$ to $(X_{2n+1} \geq X_{2n+2}) \wedge (X_{2n+2} \geq X_{2n})$.
We know that
$(X_{2n+1} \geq X_{2n+2}) \wedge (X_{2n+2} \geq X_{2n}) \rightarrow (X_{2n+1} \geq X_{2n})$
so, applying the previous theorem,
a violation counting soft constraint of $rs$ is contractible.
In Example \ref{ex:rc_alldiff},
using the natural choice of $\phi$ and $\theta$
(which maps variables $A_{i l u}$ in $\decomp(C(\vec{X}))$ to variables of the same name in $\decomp(C(\vec{X}Y))$,
constraints $\eqref{eq:diff_Ailu1}$ to themselves, and constraints $\eqref{eq:diff_Ailu2}$ to $\eqref{eq:diff_Ailu3}$),
the validity of 
$\eqref{eq:diff_Ailu3} \wedge (0 \leq A_{(n+1) l u} \leq 1) \rightarrow \eqref{eq:diff_Ailu2}$,
and the previous theorem, we establish that the violation counting soft version of $\ALLDIFF$
based on this decomposition is contractible.

There are two possible generalizations of the notion of semantic embedding that might be used
to create a broader sufficient condition for contractibility.
The first is to change the domain of $\phi$ from $(S, w)\theta$ to
an arbitrary division of $(S, w)\theta$.
The current definition essentially only applies to the division of  $(S, w)\theta$ into singletons $\{c\}$.
This generalization would allow the embedding to hold for some grouping
of constraints in the first decomposition, even when the individual constraints cannot be embedded
in the second.
A second possible generalization is to employ multiple 
pairs $\langle \phi, \theta \rangle$ with a disjunctive condition.
Such a generalization has been shown necessary to characterize conjunctive query containment/rule subsumption
when queries/rules involve pre-defined relations (i.e. constraints) \cite{Klug,lift}.
These generalizations are left for future research.

\subsection{Edit-based Violation Measures}  \label{sect:edit}

The \emph{edit-based} violation measures
use a notion of edit distance,
which is the minimum number of edit operations required to transform a word
into a word of $L_C$.
There are many possible edit operations but the common ones are:
to substitute one letter for another,
to insert a letter,
to delete a letter,
and to transpose two adjacent letters.\footnote{
Edit distance based on counting these operations is known as Damerau-Levenshtein distance.
Other well-known edit distances are defined using a subset of these operations.
}
This class includes the \emph{variable-based} violation measures \cite{softcons,warming},
since such measures are simply edit distances where substitution is the only edit operation.
The \emph{object-based} measures of \cite{costeval} are edit distances where
deletion is the only edit operation.
In \cite{warming}, an edit-based measure involving substitution, insertion and deletion is used.

To address a wide range of edit-based measures,
we generalize the measures.
We allow non-negative weights $\alpha, \beta, \gamma, \delta$ for the edit operations
substitution, insertion, deletion and transposition, respectively, and
let $n_s, n_i, n_d, n_t$ be the number of the respective operations
used in an edit.
Then we define  $m_{L}(w) = \min_{edits} \alpha n_s + \beta n_i + \gamma n_d + \delta n_t$
to be the minimum,  over all edits that transform $w$ to an element of $P(L)$, 
of the weighted sum of the edit operations.
We refer to all measures of this form as \emph{edit-based}.
Measures based on a subset of the four edit operation can be captured by giving effectively infinite weights
to the other operations.

The edit-based violation measures used for closed constraints
are not appropriate for open constraints, because they fail
to take into account that
the current sequence of variables may be extended with more variables.

For example, consider an open constraint $C$ where $L_C = abc + defghi$
and an occurrence of the constraint $C([X_1, X_2, X_3])$.
If $X_1 = d$, $X_2 = e$ and $X_3 = f$
then the unweighted edit distance of this instance to $L_C$ is 3,
even though this instance is completely accurate if the sequence of variables is extended.
Similarly, if $L_C = abc$ and 
we have an occurrence $C([X_1, X_2])$ with $X_1 = a$ and $X_2 = b$
then the unweighted edit distance is 1, even though there is no violation.

To take account of the possibility that a sequence of variables may be extended,
we employ the edit distance to $P(L_C)$, the prefix-closure of $L_C$.
In Section \ref{sect:approx} the prefix-closure was used to approximate a constraint
so that constraint propagation is sound when the constraint is open.
The use of the prefix closure here is somewhat different from its use
in that section:
rather than using $P(L_C)$ as an approximation to $L_C$,
$P(L_C)$ is used here to formulate what it means to be an (edit-based) open soft constraint.

\begin{definition}
An \emph{open edit-based violation measure} for a language $L$
is an edit-based violation measure $m_{P(L)}$ for $P(L)$.
An open edit-based violation measure $m$ for $L$ is \emph{proper} if $m(w) = 0$ iff $w \in P(L)$.
Since, in this paper, we only consider open edit-based measures
they will simply be referred to as edit-based violation measures, except in the statement of theorems.
\end{definition}

As a result of this definition,
prefix-equivalent languages have the same possible edit-based (proper) violation measures.
When $L$ is clear from the context, we simply write $m$ rather than $m_L$.

We can characterize when an open edit-based violation measure is proper.
Roughly, $m$ is improper iff some edits have zero cost and
these are able to edit some $w \in L' \backslash P(L)$ to $w' \in P(L)$.

\begin{proposition}
Let $m$ be an open edit-based violation measure for $L$ where $P(L)$ is a sublanguage of $L'$,
with weights $\alpha, \beta, \gamma$ and $\delta$.

$m$ is proper iff one of the following conditions holds:
\begin{itemize}
\item
$\min \{ \alpha, \beta, \gamma, \delta \} > 0$
\item
$\alpha = 0$, 
$\min \{ \beta, \gamma \} > 0$
and $L' \cap SameLength(P(L)) \subseteq P(L)$
\item
$\beta = 0$, 
$\min \{ \alpha, \gamma, \delta \} > 0$
and $L' \cap SubSeq(P(L)) \subseteq P(L)$
\item
$\gamma = 0$ and $L' \subseteq P(L)$
\item
$\delta = 0$, 
$\min \{ \alpha, \beta, \gamma \} > 0$
and $L' \cap Perm(P(L)) \subseteq P(L)$

\item
$\alpha = \beta = 0$, 
$\gamma > 0$
and $L' \subseteq Shorter(P(L))$
\item
$\beta = \delta = 0$, 
$\min \{\alpha, \gamma \} > 0$
and $L' \cap Subset(P(L)) \subseteq P(L)$

\end{itemize}
where, for any language $L$, \\
$SameLength(L)$ is the set of all words of the same length as a word of $L$, \\
$Shorter(L)$ is the set of all words the same length or shorter than a word of $L$, \\
$Perm(L)$ is the set of all permutations of words of $L$, \\
$SubSeq(L)$ is the set of all subsequences of a word of $L$,
and \\
$Subset(L)$ is  set of all words whose letters form a submultiset of the letters of a word of $L$.
\end{proposition}
\skipit{
\begin{proof}

Looking at the different constraints on the weights
it is easy to see that the conditions are mutually exclusive
and they cover all possible combinations of weights.
Thus to prove the characterization it is sufficient to show, in each case,
that $m$ is proper iff the remaining condition in the case holds.

If $\min \{ \alpha, \beta, \gamma, \delta \} > 0$ then
$m(w) = 0$ iff no edits are required to transform $w$ to a word of $P(L)$
iff $w \in P(L)$.
Thus, in this case, $m$ is proper.

Let $\alpha = 0$,  and $\min \{ \beta, \gamma \} > 0$.
Then, for any word $w \in L'$, $m(w) = 0$ iff 
$w$ can be edited by substitutions (and possibly transpositions if $\delta = 0$) to a word of $P(L)$
iff $w$ is the same length as a word of $P(L)$.
From the definition of proper,
$m$ is proper iff $P(L) \cap L' = SameLength(P(L)) \cap L'$,
that is $L' \cap SameLength(P(L)) \subseteq P(L)$.

Let $\beta = 0$ and $\min \{ \alpha, \gamma, \delta \} > 0$.
Then, for any word $w \in L'$, $m(w) = 0$ iff $w$ can be edited by insertions to a word of $P(L)$
iff $w$ is a subsequence of a word of $P(L)$.
Hence
$m$ is proper iff $P(L) \cap L' = Subseq(P(L)) \cap L'$.

If $\gamma = 0$ then for every word $w \in L'$, $m(w) = 0$
because $w$ can be edited by deletions to the empty word, which is in $P(L)$.
Hence
$m$ is proper iff $L' = P(L) \cap L'$, that is $L' \subseteq P(L)$.

Let $\delta = 0$ and $\min \{ \alpha, \beta, \gamma \} > 0$.
Then, for any word $w \in L'$, $m(w) = 0$ iff $w$ can be edited by transpositions to a word of $P(L)$
iff $w$ is a permutation of a word of $P(L)$.
Hence
$m$ is proper iff $P(L) \cap L' = Perm(P(L)) \cap L'$.

Let $\alpha = \beta = 0$ and $\gamma > 0$.
Then, for any word $w \in L'$, $m(w) = 0$ iff $w$ can be edited by insertions and substitutions to a word of $P(L)$ iff $w$ can be obtained by deletions and substitutions from a word of $P(L)$
iff $w$ is shorter than a word of  $P(L)$.
Hence
$m$ is proper iff $P(L) \cap L' = Shorter(P(L)) \cap L'$.

Let $\beta = \delta = 0$ and $\min \{\alpha, \gamma \} > 0$.
Then, for any word $w \in L'$, $m(w) = 0$ iff $w$ can be edited by insertions and transpositions to a word of $P(L)$ iff $w$ can be obtained by deletions and transpositions from a word of $P(L)$
iff the letters of $w$ form a submultiset of the letters of a word of $P(L)$.
Hence
$m$ is proper iff $P(L) \cap L' = Subset(P(L)) \cap L'$.
\end{proof}
}

Before presenting the main result on contractibility of edit-based soft constraints
we need to introduce some preliminary results on weighted edit distance.

We say a sequence of edit operations is in \emph{normal form} if
the edit operations are grouped by type so that
all deletions are performed before all transpositions,
which are performed before all substitutions, before all insertions,
and no letter is subject to two or more substitutions.
It is not difficult to show that any edit sequence has a corresponding sequence
in normal form that achieves the same result at lower or equal cost.

\begin{lemma}  \label{lemma:edit-normal}
Consider a weighted edit-distance and a word $\vec{a}$.
For any edit sequence that maps $\vec{a}$  to $\vec{b}$,
there is an edit sequence in normal form that also
maps $\vec{a}$  to $\vec{b}$
with a shorter or equal weighted edit distance.
\end{lemma}

It is straightforward to see that, for any edit sequence not involving transposition 
and any weighted edit measure, there is an equivalent edit sequence where
each letter is edited at most once.
Provided the edit weights satisfy a simple property, 
this result extends to edit sequences involving transposition.

\begin{proposition}  \label{prop:transp}
Consider an edit-based violation measure where
$\beta + \gamma \leq 2\delta$.
Suppose we wish to edit a word $\vec{a}$ so that it appears in a language $L$.
Then
there is an edit of minimal cost where 
no letter is subject to more than one edit operation. 
\end{proposition}
\skipit{
\begin{proof}
Suppose $\beta + \gamma \leq 2\delta$
and consider any edit sequence that maps $\vec{a}$ to $\vec{b} \in L$.
We can assume 
(Lemma \ref {lemma:edit-normal})
that edit operations are grouped:
deletions, then transpositions, substitutions, and finally insertions.

Suppose a letter $a$ that participates in a transposition also participates in another edit operation.
Then the second operation is either another transposition or a substitution.

In the former case,
consider all transposition operations that are applied to $a$.
The effect of these edits is to move $a$ from some position $i$ to a position $j$.
This sequence can be replaced by the deletion of $a$ at position $i$
and the insertion of $a$ at position $j$.
The revised edit sequence has a lower or equal cost because $\beta + \gamma \leq 2\delta$
and we assumed that at least two transpositions are involved.

In the latter case,
$aa'$ is edited to $a'a$ and later $a$ is changed to $b$, for some $a'$ and $b$.
We can achieve the same effect by substituting $a'$ for $a$ and $b$ for $a$
instead of the transposition and substitution.
The revised edit sequence has lower or equal cost if $\alpha \leq \delta$.
Alternatively,
we can replace the original edit operations by
the deletion of $a$ and the insertion of $b$ on the right of $a'$.
This revised edit sequence has lower or equal cost if $\delta \leq \alpha$,
because $\beta + \gamma \leq 2\delta \leq \alpha + \delta$.
Thus, independent of whether $\alpha \leq \delta$ or $\delta \leq \alpha$,
a lower cost edit sequence is obtained with fewer instances of a letter
involved in two edit operations.

The remaining possibility is that a substitution operation is applied twice to a letter.
It is clear that the first substitution operation can be omitted.

Repeatedly applying normal form transformations and the edit modifications
described above, all occurrences of a letter being edited twice
can be removed.
\end{proof}
}

In particular, this lemma holds when the edit operations are unweighted
(that is, when $\alpha = \beta = \gamma = \delta$).
The property that each letter is edited at most once is important for 
network flow implementations of propagators
such as the propagators for soft $\REGULAR$ in \cite{warming,open3}.

Edit-based violation measures are monotonic with respect to both the weights and the language.

\begin{lemma}  \label{lemma:mono1}
Let $m$ ($m'$) be edit-based violation measures with weights $\alpha, \beta, \gamma, \delta$
(respectively $\alpha', \beta', \gamma', \delta'$)
for the same language.
If $\alpha \leq \alpha'$, $\beta \leq \beta'$, $\gamma \leq \gamma'$ and $\delta \leq \delta'$ then,
for all words $w$,
$m(w) \leq m'(w)$.
\end {lemma}
\skipit{
\begin{proof}
For every word $w$,
consider an edit that achieves the minimum violation $m'(w)$.
Let $n_s, n_i, n_d, n_t$ be the number of the respective operations
used in the edit.
Then
$m'(w) = \alpha' n_s + \beta' n_i + \gamma' n_d + \delta' n_t
\geq  \alpha n_s + \beta n_i + \gamma n_d + \delta n_t
\geq m(w)$.
Hence $m(w) \leq m'(w)$.
\end{proof}
}

\begin{lemma}  \label{lemma:mono2}
Let $m_1$ and $m_2$ be edit-based violation measures 
with the same weights, 
for languages $L_1$ and $L_2$ respectively.
If $L_1 \subseteq L_2$ then
for all words $w$,
$m_1(w) \geq m_2(w)$.
\end {lemma}
\skipit{
\begin{proof}
For every word $w$,
any edit to $L_1$ is also an edit to $L_2$.
Since an edit-based violation measure minimizes over all edits,
we must have $m_1(w) \geq m_2(w)$.
\end{proof}
}

In many cases, edit-based violation measures lead to contractible soft constraints.

\begin{theorem}  \label{thm:edit}
Let $C_s$ be a soft constraint with an open edit-based violation measure, and suppose
$\min \{ \alpha, \beta, \gamma \} \leq \delta$.

Then $C_s$ is contractible.
\end{theorem}
\skipit{
\begin{proof}
Consider the sequence of edits that transforms an instance  $\vec{a}a'$ of $\vec{X}Y$
into an element $\vec{b}$ of $P(L_C)$ at minimum cost.
By Lemma \ref{lemma:edit-normal} we can assume that all deletions occur before any transpositions,
and all insertions and substitutions occur after all transpositions.
We now identify modifications of this sequence of edits 
that transform $\vec{a}$ into an element of $P(L_C)$
at lower (or equal) cost  than the original sequence.

If $a'$ is deleted in the original sequence, then the sequence of edits omitting this deletion
transforms $\vec{a}$ to $\vec{b}$ at lower or equal cost.
Otherwise, if $a'$ is not involved in a transposition, 
then the subsequence of edits that do not involve $a'$
transforms $\vec{a}$ into a prefix of $\vec{b}$ (which is an element of $P(L_C)$).
The subsequence has a lower or equal cost, since it involves a subset of the edits.

The remaining possibility is that $a'$ is involved in a transposition.
Let $p$ be the position of $a'$ after all transpositions.
The sequence of edits that omits all transpositions involving $a'$ and then
inserts $a'$ at position $p$ transforms $\vec{a}$ to $\vec{b}$.
These edits have a lower or equal cost if $\beta \leq \delta$.

Alternatively, let the length of $\vec{a}a'$ after all deletions be $n+1$
(so that $a'$ is in position $n+1$).
Every transposition involving position $n+1$ in the original sequence
can be replaced by a substitution that replaces the letter at position $n$
by the letter at position $n+1$ at the corresponding stage of the original transformation.
This transforms $\vec{a}$ into a prefix of $\vec{b}$
at  lower or equal cost if $\alpha \leq \delta$.

Finally,
let $\vec{a_1}a'\vec{a_2}$ be the result of deletions and transpositions on $\vec{a}a'$.
The length of $\vec{a_2}$
is a lower bound for number of transpositions involving $a'$ in editing $\vec{a}$ into $\vec{b}$.
The sequence of edits that deletes all letters of $\vec{a_2}$
and applies all substitutions and insertions that apply to $\vec{a_1}$
transforms $\vec{a}$ into a prefix of $\vec{b}$.
These edits have a lower or equal cost if $\gamma \leq \delta$
since transpositions are replaced by deletions and some edits might now be omitted.

In each case, for all words $\vec{a}a'$, we find that  $\vec{a}$ has a smaller weighted edit distance
to $P(L_C)$ than $\vec{a}a'$.
This demonstrates that the violation measure is non-decreasing and
hence, by Proposition \ref{prop:class}, $C_s$ is contractible.
\end{proof}
}

Example \ref {ex:notcont} below shows that this theorem cannot be strengthened
without imposing extra conditions on $C_s$.

It follows from the theorem that 
edit-based measures that only involve substitutions, insertions and deletions
provide contractible constraints.
Thus the variable-based measures \cite{softcons,warming},
the object-based measures \cite{costeval},
and the edit-based measures of  \cite{warming}
induce contractible soft constraints. 

For order-free constraints, transposition is not needed in an edit
and can be effectively given infinite weight.  Thus, by Theorem \ref{thm:edit}, we have

\begin{corollary}  \label{cor:order-free}
If $C$ is an order-free constraint and the corresponding soft constraint $C_s$ 
is based on an open edit-based violation measure $m$,
then $C_s$ is contractible.
\end {corollary}

We also have the following curious result.

\begin{corollary}   \label{cor:zero}
Let $C_s$ be a soft constraint based on an open edit-based violation measure $m$ 
with weights $\alpha, \beta, \gamma, \delta$
for the hard constraint $C$.
If any of $\alpha, \beta, \gamma$, or $\delta$ is 0
then $C_s$ is contractible.
\end {corollary}
\skipit{
\begin{proof}
If $\alpha$, $\beta$ or $\gamma$ is 0 then
the condition of Theorem \ref{thm:edit} is satisfied and consequently $C_s$ is contractible.
If $\delta$ is 0 then transpositions can place the letters in a word in any order, at no cost.
Let
\[
C'([X_1, \ldots, X_n]) \leftrightarrow \bigvee_\pi C([X_{\pi(1)}, \ldots, X_{\pi(n)}])
\]
where the disjunction is over all permutations $\pi$ of $1..n$.
Then the violation measure $m$ of $C$ is equal to the violation measure $m'$ of $C'$,
where $m'$ uses the same weights as $m$.
$C'$ is order-free and, by Corollary \ref{cor:order-free}, is contractible.
\end{proof}
}

From these results we see that soft constraints based on a wide range of edit-based measures
are contractible.
However, when transpositions are allowed and have a comparatively low cost,
an edit-based violation measure
can lead to a soft constraint that is not contractible.

\begin{example}  \label{ex:notcont}
Consider a constraint $C$ with $L_C = (ab)^* + (ab)^*a$,
which is a prefix-closed language,
and consider the corresponding soft constraint $C_s$ that uses an edit-based violation measure.
Suppose $\delta < \min\{\alpha, \beta, \gamma\}$.
The word $abba$ has edit distance $\delta$, by transposing the last two letters,
but its prefix $abb$ has edit distance $\min \{\alpha, \beta, \gamma\}$,
since we could either
substitute $a$ for $b$,
insert $a$ before the second $b$, 
or delete a $b$.
Thus the weighted edit-based violation measure is not non-decreasing and hence,
by Proposition \ref{prop:class},
$C_s$ is not contractible.
\end{example}

This example reinforces a point made earlier: the introduction of $P(L_C)$ to the
definition of edit-based violation measure plays a different role than its use for hard constraints;
in this case, its use does not ensure contractibility.

\section{Contractible Approximations of Soft Constraints}  \label{sect:softapprox}

Although we have identified powerful sufficient conditions for
soft constraints to be contractible,
we must also be able to support uncontractible soft constraints.
As with hard constraints, when a soft constraint is uncontractible
we can use a contractible approximation as the basis for filtering
while the constraint is open.

We reformulate the notion of tight approximation for soft constraints
of the form $m(\vec{X}) \leq Z$ as follows.
A violation measure $m_1$ is an \emph{approximation} of the violation measure $m$
if, for all words $\vec{a}$, $m_1(\vec{a}) \leq m(\vec{a})$.
We order violation measures with the pointwise extension of the ordering on the reals:
$m_1 \leq m_2$ iff $\forall \vec{a} ~ m_1(\vec{a}) \leq m_2(\vec{a})$.
A contractible approximation $m_1$ to a violation measure $m$  is \emph{tight} if,
for all contractible functions $m_2$, if $m_1 \leq m_2 \leq m$ then $m_2 = m_1$.
Given two contractible approximations $m_1$ and $m_2$ to a violation measure $m$,
we say $m_2$ is \emph{tighter} than $m_1$ if $m_1 \leq m_2$.
We write $m^*$ to denote the tightest contractible approximation of $m$.

We can characterize the tightest contractible approximation of a violation measure,
independent of how the violation measure is formulated.

\begin{proposition}   \label{prop:m*}
Let $m$ be a violation measure.
The tightest contractible approximation to $m$ is characterized by
$m^*(\vec{a}) = \inf_{\vec{b}} m(\vec{a}\vec{b})$,
where the infimum is taken over all finite sequences ${\vec{b}}$.
\end {proposition}
\skipit{
\begin{proof}
By definition, $m^*(\vec{a}) \leq m(\vec{a})$, so $m^*$ approximates $m$.
Consider a sequence $\vec{a}$ and a letter $c$.
$m^*(\vec{a}c) = \inf_{\vec{b}} m(\vec{a}c\vec{b})  \geq \inf_{c\vec{b}} m(\vec{a}c\vec{b}) 
\geq \inf \{ m(\vec{a}), \inf_{c\vec{b}} m(\vec{a}c\vec{b}) \} = m^*(\vec{a})$.
Thus $m^*$ is contractible.

Suppose some function $k$ is a strictly tighter contractible approximation than $m^*$.
Then, for some $\vec{a}$, $k(\vec{a}) > m^*(\vec{a})$, that is,
$k(\vec{a}) >  \inf_{\vec{b}} m(\vec{a}\vec{b})$.
Hence, there is a $\vec{d}$ such that $k(\vec{a}) > m(\vec{a}\vec{d})$.
But, for any $\vec{c}$,
$k(\vec{a}\vec{c})  \geq k(\vec{a})$.
Thus we have $m(\vec{a}\vec{d}) > m(\vec{a}\vec{d})$.
This contradiction shows that $k$ cannot exist; $m^*$ is the tightest contractible approximation to $m$.
\end{proof}
}

This proposition only provides a mathematical characterization;
it does not suggest an implementation.
Indeed, it appears very difficult to implement this tightest contractible approximation, in general,
in contrast to the tightest contractible approximation of hard constraints.
Nevertheless, we can identify some contractible approximations.

\subsection{Decomposition-based Violation Measures}

One way to obtain a contractible approximation 
to a decomposition-based soft constraint
is to ignore parts of a decomposition that cause incontractibility.
A \emph{weakening} of a decomposition of a constraint $C(\vec{X})$ is a function
that, for every sequence $\vec{X}$, 
maps the decomposition $(\vec{X}, \vec{U}, T, S, w)$
to $(\vec{X}, \vec{U}, T, S', w')$
where $(S', w')$ is a sub-weighted set of $(S, w)$.
For this weakened decomposition we can apply the sufficient condition of Theorem \ref{thm:decomp}.

\begin{proposition}  \label{prop:approx_decomp}
Consider a decomposition-based violation measure $m$ for a constraint $C(\vec{X})$
and a weakening $W$ of the decomposition.
Suppose $m$ is defined via a monotonic combining function.
If, for every sequence  $\vec{X}$,
the weakening of the decomposition of $C(\vec{X})$ can be semantically embedded in
the weakening of the decomposition of $C(\vec{X}Y)$
then the measure $m'$ defined by using the weakened decompositions
is a contractible approximation of $m$.
\end{proposition}
\skipit{
\begin{proof}
$m'$ is an approximation of $m$ because the combining function is monotonic
and the weakened decomposition employs a sub-weighted set of the original decomposition.
$m'$ is contractible by application of Theorem \ref{thm:decomp}.
\end{proof}
}

This result shows an approach to finding a contractible approximation to $C_s(\vec{X})$.
However, there is no guarantee that it will find a good approximation;
in the worst case it might provide only the trivial approximation, where all of $C(\vec{X})$ is ignored.
Nevertheless, it appears to be useful.

The next example presents an uncontractible decomposition-based soft constraint.
It employs
a decomposition of the global cardinality constraint $\GCC$ given in \cite{decomp}.

\begin{example}  \label{ex:GCC}
Consider the global cardinality constraint $\GCC(\vec{X}, \vec{l}, \vec{u})$ 
with type $T$ that maps each $X_i$ to $1..d$,
which we denote by $\GCC_T$.
This constraint expresses that,
for each value $t$ in $1..d$,
the number of occurrences of $t$ in $\vec{X}$ lies between $l_t$ and $u_t$
($u_t$ may be infinite).
$\vec{l}$ and $\vec{u}$ are fixed.
To define the decomposition of  \cite{decomp}
we need to introduce variables $A_{i l u}$ of type $\{0, 1\}$
and $N_{l u}$ of type non-negative integers,
and elementary constraints as follows.
Let $n = |\vec{X}|$.

For $1 \leq i \leq n$, $1 \leq l \leq u \leq d$ and $1 \leq k < u$
we have the constraints
\begin{equation}  \label{eq:gcc_Ailu1}
A_{i l u} = 1 \leftrightarrow X_i \in [l, u]
\end{equation}

\begin{equation}  \label{eq:gcc_Nlu2}
N_{l u} = \sum_{i=1}^n A_{i l u}
\end{equation}

\begin{equation}  \label{eq:gcc_Nlu3}
N_{1 u} = N_{1 k} + N_{(k+1) u} 
\end{equation}

\begin{equation}  \label{eq:gcc_Nlu4}
\sum_{j=l}^u  l_j \leq N_{l u} \leq \sum_{j=l}^u  u_j 
\end{equation}

\noindent
Formally, the decomposition of $\GCC_T$ is
$(\vec{X}, \vec{A}, \vec{N}, T', S, w)$
where $T'$ is the extension of $T$ to $\vec{A}$ and $\vec{N}$,
$S$ is the collection of
$\eqref{eq:gcc_Ailu1}$, $\eqref{eq:gcc_Nlu2}$, $\eqref{eq:gcc_Nlu3}$, and $\eqref{eq:gcc_Nlu4}$,
and $w$ is a constant function.
It is easy to establish that
$\GCC_T(\vec{X}, \vec{l}, \vec{u}) \leftrightarrow \exists \vec{A} \in T'(\vec{A}) \ \exists \vec{N} \in T'(\vec{N}) ~ S$.

When $\vec{X}$ is extended by $Y$, the decomposition contains extra variables
$A_{(n+1) l u}$,
extra constraints of type \eqref{eq:gcc_Ailu1} involving $Y$ and the new variables,
and replaces constraints \eqref{eq:gcc_Nlu2}  by

\begin{equation}  \label{eq:gcc_Nlu5}
N_{l u} = \sum_{i=1}^{n+1} A_{i l u}
\end{equation}

Consider an occurrence of the constraint
$\GCC_T([X_1, X_2], [0,1,0,0], [2,2,2,2])$
where $T(X_i)$ is $1..4$.
Consider a valuation $v$ where $X_1 = 1, X_2 = 1$.
For all extensions of $v$ to $\vec{A}$ and $\vec{N}$
there will be an elementary constraint violated
(fundamentally because the lower bound for occurrences of the domain value 2 has not been satisfied).
If $\vec{X}$ is extended by $X_3$ and $v$ has $X_3 = 2$
then $v$ can be extended to $\vec{A}$ and $\vec{N}$ in the obvious way
to satisfy all elementary constraints.
Thus any proper violation measure for $\GCC$
based on this decomposition
is not contractible.

Let $m$ be a proper violation measure that is defined with a combining function
that is monotonic and has unit 0.
If we weaken the decomposition by ignoring the lower bounds in \eqref{eq:gcc_Nlu4}
then we have a contractible approximation $m'$ of $m$.
(This is essentially the same as for the tight contractible approximation of the hard $\GCC$ constraint,
which is also obtained by ignoring lower bounds.
This point is not so surprising when we recall that the hard constraint is a special case of the soft constraint.)
We can see this using the natural semantic embedding 
(which maps all constraints to themselves, except that $\eqref{eq:gcc_Nlu2}$ is mapped to $\eqref{eq:gcc_Nlu5}$)
and Proposition \ref{prop:approx_decomp}.
\end{example}

We conjecture that the weakening of the soft $\GCC$ constraint in this example
is its tightest contractible approximation.
However, the many variables and constraints in the decomposition
make it difficult to confirm this conjecture.

\subsection{Edit-based Violation Measures}

Recall that an edit-based violation measure $m$ is contractible if
$\delta \geq \min\{\alpha, \beta, \gamma\}$ (Theorem \ref{thm:edit}).
If $\delta < \min\{\alpha, \beta, \gamma\}$ then $m$ might be uncontractible
and we must consider contractible approximations.
We can provide generic contractible approximations for edit-based soft constraints
by modifying the weights to accord with the sufficient conditions of Theorem \ref{thm:edit}
and Corollary \ref{cor:zero}.

\begin{proposition}  \label {prop:approxedit}
Let $m$ be an open edit-based violation measure for a constraint $C$ 
with weights $\alpha, \beta, \gamma, \delta$ where $\delta < \min\{\alpha, \beta, \gamma\}$.
Then the following violation measures are contractible approximations of $m$ for $C$.

\begin{enumerate}
\item
$m_1$ based on weights $\delta, \beta, \gamma, \delta$  (that is, $\alpha := \delta$)
\item
$m_2$ based on weights $\alpha, \delta, \gamma, \delta$   (that is, $\beta := \delta$)
\item
$m_3$ based on weights $\alpha, \beta, \delta, \delta$   (that is, $\gamma := \delta$)
\item
$m_4$ based on weights $\alpha, \beta, \gamma, 0$ (that is, $\delta := 0$)
\item
$m_5$ defined by $m_5(w) = \max \{m_1(w), m_2(w), m_3(w), m_4(w)\}$
\end{enumerate}
\end {proposition}
\skipit{
\begin{proof}
By Lemma \ref{lemma:mono1}, for any $w$, $m_1(w) \leq m(w)$,
$m_2(w) \leq m(w)$, $m_3(w) \leq m(w)$, and $m_4(w) \leq m(w)$.
It then follows from the definition of $m_5$ that $m_5(w) \leq m(w)$.
Thus $m_1$, $m_2$, $m_3$, $m_4$ and $m_5$ are approximations of $m$.
By Theorem \ref{thm:edit}, $m_1$, $m_2$, and $m_3$ are contractible
and, by Corollary \ref{cor:zero}, $m_4$ is contractible.
For any word $w$ and letter $a$,
\[
\begin{array}{rcl}
m_5(wa) & =     & \max \{m_1(wa), m_2(wa), m_3(wa), m_4(wa)\}  \\
               & \geq & \max \{m_1(w), m_2(w), m_3(w), m_4(w)\} \\
               & = & m_5(w)
\end{array}
\]
using the contractibility of $m_1, \ldots, m_4$.

Thus $m_5$ is contractible.
\end{proof}
} 

Note that, by Lemma \ref{lemma:mono1}, other uses of Corollary \ref{cor:zero} 
yield only measures that are not as tight as $m_1$, $m_2$, or $m_3$.
Clearly $m_5$ is the tightest of these approximations.
However, in general, this approximation is not tight,
as the following example shows.

\begin{example}  \label{ex:m4}
Let $L = (abc)^*$, so that $P(L) = L \cup La \cup Lab$.
Let $\alpha = \beta = \gamma = 4$ and $\delta = 1$.
Consider $w = bbb(abc)^3ca$.
Two kinds of edits are needed, addressing the initial $b$'s and the trailing $ca$.
Then $m(w) = 12$ from substituting for the first and third $b$, and deleting the last $c$.
$m(wb) = 10$ using the same substitutions and two transpositions on $c$.
Thus $m$ is not contractible.

Notice that the initial $b$'s in $w$ are too far from the end of $w$ to be cheaply addressed
by transpositions.  
For example, the cost of moving the third $b$ to the trailing $ca$  
is 6, which is more expensive than addressing it by substitution.
The other $b$'s are even more expensive to address by transposition.
Thus the minimal cost of addressing the initial $b$'s is 8.
The minimal cost of addressing the trailing $ca$ arises when a $b$ is appended to the end of $w$
and $c$ is transposed twice.
This has a cost of 2, and it is easy to see that no word appended to $w$ will allow
$ca$ to be addressed by a single transposition.
Thus the tightest approximation to $m$ has $m^*(w) = 10$.

Now consider the approximations in Proposition \ref{prop:approxedit}.
If we reduce $\alpha$ to 1 then $m_1(w) = 4$ by applying four substitutions.
If we reduce $\beta$ to 1 then $m_2(w) = 8$ by inserting $a$ and $c$ around each initial $b$
and inserting $ab$ before the last $c$.
If we reduce $\gamma$ to 1 then $m_3(w) = 4$ by deleting the three $b$'s and the last $c$.
If we reduce $\delta$ to 0 then $m_4(w) = 4$ by applying transpositions to reorder $w$ to $(abc)^4bb$
and then substituting $a$ for $b$.
Thus $m_5(w) = 8$.

This shows that $m_5$ is not the tightest contractible approximation to $m$, since $m_5(w) \neq m^*(w)$.
\end{example}

The question now arises: how to express $m^*$ in edit-based terms
so that a closed propagator for $m(\vec{X}) \leq Z$ might be adapted to implement
$m^*(\vec{X}) \leq Z$, as was done for hard constraints in Section \ref{sect:approx}.
Disappointingly, this turns out to be impossible, in general.

We first establish a straightforward lemma that gives a simple way of identifying the value of $m^*(w)$
in some cases.
\begin{lemma}  \label{lemma:m*}
Let $m^*$ be the tightest contractible approximation to an edit-based violation measure $m$.
Let $w$ be a word.
If, for all words $u$, $m(wu) \geq m(w)$ then $m^*(w) = m(w)$.
\end{lemma}
\skipit{
\begin{proof}
If, for all words $u$, $m(wu) \geq m(w)$ then $\inf_u m(wu) \geq m(w)$.
Thus $m^*(w) \geq m(w)$.
Since $m^*$ approximates $m$, $m^*(w) \leq m(w)$.
Hence, $m^*(w) = m(w)$.
\end{proof}
}

Now we show that,
in general, 
the tightest contractible approximation $m^*$ to an edit-based violation measure $m$
cannot be expressed as a proper edit-based violation measure.

\begin{theorem}  \label{thm:inexpressive}
There is an open edit-based violation measure $m$ for a language $L$ such that
its tightest contractible approximation cannot be expressed as 
a proper edit-based violation measure 
on any language.
\end{theorem}
\skipit{
\begin{proof}
Consider the alphabet $\Sigma = \{a, b, c, d\}$.
As in Example \ref{ex:m4}, 
let $L = (abc)^*$ (so $P(L) = L \cup La \cup Lab$), 
consider $P(L)$ as a sublanguage of $\Sigma^*$,
and let $m$ be the edit-based violation measure for $L$ where
$\alpha = \beta = \gamma = 4$ and $\delta = 1$.
As shown in Example \ref{ex:m4}, $m$ is not contractible.
Note that $m$ is proper.
Let $m^*$ be the tightest contractible approximation to $m$.
Suppose $m^*$ can be expressed as a proper edit-based violation measure $m'$ on some language $L'$.

Suppose there is some word $w$ such that $w \in L'\backslash P(L)$.
Then $m'(w) = 0$.
Hence $m^*(w) = 0$ and, from Proposition \ref{prop:m*},
there is a word $u$ such that $m(wu) = 0$.
Since $m$ is proper, $wu \in P(L)$ and hence $w \in P(L)$.
This contradiction shows that no such $w$ exists
and hence $L' \subseteq P(L)$.

For every $w \in L$, $m(w) = 0$.  Hence $m^*(w) = 0$ and $m'(w) = 0$.
Hence $w \in L'$, since $m'$ is proper.  Hence $L \subseteq L'$.

There are weights $\alpha', \beta', \gamma'$, and $\delta'$ used to define $m'$.
We now consider different words $w$ of $\Sigma^*$ and derive conditions on the weights of $m'$.
We use the fact that every edit of $w$ to $P(L)$ must have cost greater than or equal to $m^*(w)$.
Because $L' \subseteq P(L)$, 
the conditions we derive about editing a word to $P(L)$ also apply to $L'$.

$w = d$. \\
$m(w) = 4$ by deleting $d$ and no word $wu$ has a smaller violation measure
because $d$ must be deleted or substituted.  
Thus, by Lemma \ref{lemma:m*}, $m^*(w) = 4$.
$w$ might be edited to $L$ (and hence also $L'$)
by deleting $d$ or substituting $a$ for $d$.
This gives rise to the conditions $\gamma' \geq 4$ and $\alpha'  \geq 4$ since, for example,
if $\gamma' = 3$ then $m'(w) = 3 \neq m^*(w)$.

$w = bc(abc)^3$. \\
$m(w) = 4$, by inserting $a$ at the beginning of $w$.
No word $wu$ has a lower cost because the initial $bc$ is too far from the end of $w$ to
use transposition from $u$ at a lower cost.  Thus $m^*(w) = 4$.
From this we obtain the condition $\beta' \geq 4$, among others.

$w = ba$. \\
$m(w) = 1$ by transposition and we find that $m^*(w) = 1$.
Because we know that $\alpha' \geq 4$, $\beta' \geq 4$ and $\gamma' \geq 4$,
we must have $\delta' = 1$.

[We can now establish that $L' \subseteq (a+b+c)^*$.
For any word $w$ involving $d$, $m(w) \geq 4$, since the $d$ must be deleted or substituted.
That includes $wu$, for any $u$, and hence $m^*(w) \geq 4$ for any word containing $d$.
Consequently, also $m'(w) \geq 4$ and, since $m'$ is proper,
$L'$ does not contain a word involving $d$.]

$w = adc$. \\
$m(w) = 4 = m^*(w)$.
Given that $\delta' = 1$ and the small size of $w$,
no word is edit distance 4 from $w$ using transposition alone.
But $\alpha' \geq 4$, $\beta' \geq 4$ and $\gamma' \geq 4$,
so the minimum cost edit from $w$ to $L'$ does not involve transposition
and the edit consists of a single operation.
Since $L'$ does not contain words involving $d$,
the only candidates are deletion or substitution of $d$.
The deletion results in $ac$ which is not in $L'$ since it is not in $P(L)$.
Hence
the only edit that can achieve this cost is a substitution of $b$ for $d$,
and $\alpha' = 4$.

$w = d(abc)^3$.  \\
As with $w = d$, $m^*(w) = m(w) = 4$.
Given that  $\alpha' \geq 4$, only deletion of $d$ can achieve this cost.
Hence $\gamma' = 4$.

$w = bc(abc)^3$, again. \\
Given that  $\alpha' \geq 4$ and $\gamma' \geq 4$,
the only edit that can achieve $m^*(w) = 4$ is an insertion.
Hence $\beta' = 4$.

Thus the weights for $m'$ are exactly the same as the weights for $m$.
For every word $w$, $m'(w) \geq m(w)$, by Lemma \ref{lemma:mono2}.
From the definition of $m^*$, $m(w) \geq m^*(w)$.
But $m' = m^*$, by assumption, and hence $m^* = m$.
But this is a contradiction, because by definition $m^*$ is  contractible, while $m$ is not.
Thus the assumption that $m^*$ can be expressed as an edit-based violation measure is false.
\end{proof}
}

The language and violation measure demonstrating this claim are those from Example \ref{ex:m4}.
Given that the language is so simple,
we can expect that many uncontractible edit-based violation measures
cannot be tightly approximated by a contractible edit-based violation measure.
This contrasts markedly with our work on hard constraints in Section \ref{sect:approx},
where tight contractible approximations of several uncontractible hard constraints
were formulated in terms of the original hard constraint.

It suggests some difficulties in implementing tight contractible approximations.
It seems that the edit-based implementation of the closed constraint
is not a suitable basis for implementing the tight approximation.
At least we need a different framework if we are to have a comprehensive method
to derive open D-consistent propagators for incontractible soft constraints.

It is demonstrated in \cite{badSR} that using an approximation of the violation measure
of a \emph{closed} edit-based soft constraint can lead to incorrect answers to constraint problems.
However, using a non-tight contractible approximation in an \emph{open} constraint is less serious,
assuming the correct violation measure is used for the closed constraint:
the search may perform less-then-optimal pruning, leading to a greater search space than for
a tight contractible approximation, but not to incorrect answers.
Thus, Theorem \ref{thm:inexpressive} does not represent a failure of correctness,
only a degree of inefficiency if a non-tight edit-based contractible approximation is used.

\section{Discussion}   \label{sect:disc}

We have discussed open constraints where variables are added to the right-hand end of the sequence.
This directly affects the characterization of contractibility and the definition of open D-consistency.
If, instead, variables are added to the left-hand end then
the appropriate characterization of contractibility is suffix-closure.
If additions may be made at either end then contractibility requires both closures,
which corresponds to closure under taking subwords.\footnote{
A word $w$ is a \emph{subword} of $a_1 \ldots a_n$ if
$w$ is empty or
has the form $a_{i} a_{i+1} \ldots a_{j}$ for some $1 \leq i \leq j \leq n$.
}
Constraints like $\SEQUENCE$ and $\CONTIG$ are subword-closed.
Of course, all order-free constraints are subword-closed.
On the other hand, the lexicographic ordering constraint $\leq_{lex}$ and 
the precedence constraint $s \prec_{\vec{X}} t$ are prefix-closed but not suffix-closed, that is,
they are contractible if variables are added on the right,
but not if variables are added on the left.

If additional variables may be inserted anywhere within the sequence then
contractibility corresponds to closure under taking subsequences.\footnote{
A word $w$ is a \emph{subsequence} of $a_1 \ldots a_n$ if
$w$ is empty or
has the form $a_{i_1} \ldots a_{i_k}$ for some $k \leq n$ where
$1 \leq i_1 < i_2 < \cdots < i_k \leq n$.
}
Apart from the order-free constraints,
it is not clear whether there is any useful constraint that is closed under taking subsequences.

In \cite{open2} a dual notion to contractibility, called extensibility, is investigated.
In contrast to contractibility, in general there is no closure operation corresponding to extensibility
and consequently no tightest extensible approximation.

We have seen some differences between contractibility for hard and soft constraints.
For hard constraints contractibility depends on the relation whereas
for soft constraints it depends on the violation measure.
For example, the soft $\REGULAR$ constraint is contractible under the edit-based measure of \cite{warming}
but not under decomposition-based measures.
We have also seen that many tight contractible approximations of hard constraints
are similar to, though weaker than, the hard constraint.
On the other hand, for many soft constraints it appears that the tight contractible approximations
cannot be expressed in the same way as the soft constraint.
This suggests that it may be difficult to formulate full open D-consistent propagators, for example,
for uncontractible open soft constraints.

There are several similarities between violation measures and other treatments of
soft constraints.
For example, 
the Valued CSP \cite{VCSP} and the Semi-Ring CSP \cite{semiring} frameworks  
define a soft constraint essentially as a function from valuations to an ordered set
(the set may be partially ordered in the case of SCSPs) that might be considered a violation measure.
Both frameworks use a combining function to extend this definition to a collection of constraints,
and so they are, in many ways, like decomposition-based violation measures.
However both frameworks consider only closed constraints and focus on 
finite relations defined extensionally.

Weighted violation measures are used in \cite{MBL07,MBL09}.
As noted earlier, the decomposition measures presented here generalize
the weighted decomposition measures for $\Sigma$-$\ALLDIFF$ and $\Sigma$-$\GCC$ \cite{MBL07,MBL09}.
However, the edit-based violation measures presented here
do not generalize 
the weighted edit distance for $\Sigma$-$\ALLDIFF$
and $\Sigma$-$\REGULAR$
of \cite{MBL07,MBL09}.
These measure use only substitution edits but they
assign weights to each variable.

Violation measures play a similar role to query measures \cite{MS89} 
that were used to specify preferences on query solutions in a CLP system.
In this context, contractible violation measures are similar to pruning measures in \cite{MS89}
in that both are non-decreasing functions, although over different domains,
and both permit the safe pruning of search trees.

Contractible global soft constraints are amenable to a nested representation \cite{BBGM14}
in a distributed constraint optimization setting, which has significant performance gains
over other representations \cite{BBGM14}.

Finally, we note that
the semantics of soft constraints are examples of quantitative languages,
in the terminology of \cite{quantlang}.
From this point of view,
an approximation of a violation measure is a  quantitative language inclusion.
However, \cite{quantlang} focuses on languages of infinite words defined via automata,
so the results of \cite{quantlang} do not seem to have application to the subject of this paper.
In \cite{Colcombet} a notion of cost function on languages of finite words is used
but this is only used to define equivalence classes and is not related to this paper.

\section{Conclusions}

We have introduced the notion of contractibiliity of global constraints,
which ensures that constraint propagation for closed constraints is safe for open constraints,
and characterized it in language-theoretic terms.
The concept of contractibility is remarkably robust.
It is based only on the relation, or language, defining the constraint.
Thus it is independent of the form of propagator used (monolithic or decomposed)
and the consistency condition (if any) that characterizes the propagation.

Contractibility appears to be central to the re-use of closed constraint propagators for open propagation.
When a constraint is contractible we only need to modify a closed propagator
to support the addition of variables.
When a constraint is incontractible we also need a contractible approximation of the propagator,
for use while the constraint is open, in addition to the closed propagator.
We showed that the use of a tight contractible approximation and domain consistent closed propagators
achieves open D-consistency of the resulting open propagator.
Furthermore, for many hard constraints 
($\REGULAR$, $\CFG$, $\GCC$, and many others)
we showed that the tightest contractible approximation has a similar form to the original constraint,
and hence can be propagated by the same techniques.
This suggests that a close integration of the two propagation phases will be easy for these constraints.

To address soft constraints, we formulated two general classes of soft constraints that include
most previous proposals of soft constraints.
For the two classes -- based on decomposition and edit-distance, respectively --
we  identified properties and developed mathematical tools for reasoning about them,
which we used to demonstrate the contractibility of a wide range of soft constraints.
We identified pragmatic contractible approximations of soft constraints in these classes.
However, we also established that the tightest contractible approximation of an edit-based soft constraint
is not expressible, in general, as an edit-based constraint.
This suggests difficulties in designing open D-consistency propagators in the general case
but fortunately many edit-based soft constraints are contractible.

These results provide a good basis for adapting existing algorithms and implementations
of global constraint propagators to open constraints.

\subsubsection*{Acknowledgements}
Thanks to the referees of
this paper and
previous conference papers, whose thorough reviews and detailed comments improved this paper.
The work in this paper was mostly conducted while the author was employed by NICTA.

\bibliographystyle{acmtrans}
\bibliography{entire_final}

\begin{thebibliography}{}

\bibitem[\protect\citeauthoryear{Bart{\'{a}}k}{Bart{\'{a}}k}{1999}]{Bartak99}
{\sc Bart{\'{a}}k, R.} 1999.
\newblock Dynamic constraint models for planning and scheduling problems.
\newblock In {\em New Trends in Contraints}. 237--255.

\bibitem[\protect\citeauthoryear{Bart{\'{a}}k}{Bart{\'{a}}k}{2003}]{dynamic}
{\sc Bart{\'{a}}k, R.} 2003.
\newblock Dynamic global constraints in backtracking based environments.
\newblock {\em Annals {OR}\/}~{\em 118,\/}~1-4, 101--119.

\bibitem[\protect\citeauthoryear{Beldiceanu and Carlsson}{Beldiceanu and
  Carlsson}{2001}]{cardpath}
{\sc Beldiceanu, N.} {\sc and} {\sc Carlsson, M.} 2001.
\newblock Revisiting the cardinality operator and introducing the
  cardinality-path constraint family.
\newblock In {\em Logic Programming, 17th International Conference, {ICLP}
  2001, Paphos, Cyprus, November 26 - December 1, 2001, Proceedings}. 59--73.

\bibitem[\protect\citeauthoryear{Beldiceanu, Carlsson, and Rampon}{Beldiceanu
  et~al\mbox{.}}{2005}]{GCcatalog}
{\sc Beldiceanu, N.}, {\sc Carlsson, M.}, {\sc and} {\sc Rampon, J.-X.} 2005.
\newblock Global constraint catalog.
\newblock Tech. Rep. T2005:08, SICS.
\newblock Current version available at {\small
  \verb"http://sofdem.github.io/gccat/"}.

\bibitem[\protect\citeauthoryear{Beldiceanu and Contejean}{Beldiceanu and
  Contejean}{1994}]{BC}
{\sc Beldiceanu, N.} {\sc and} {\sc Contejean, E.} 1994.
\newblock Introducing global constraints in {CHIP}.
\newblock {\em Mathematical Computer Modelling\/}~{\em 20,\/}~12, 97--123.

\bibitem[\protect\citeauthoryear{Beldiceanu and Petit}{Beldiceanu and
  Petit}{2004}]{costeval}
{\sc Beldiceanu, N.} {\sc and} {\sc Petit, T.} 2004.
\newblock Cost evaluation of soft global constraints.
\newblock In {\em Integration of {AI} and {OR} Techniques in Constraint
  Programming for Combinatorial Optimization Problems, First International
  Conference, {CPAIOR} 2004, Nice, France, April 20-22, 2004, Proceedings}.
  80--95.

\bibitem[\protect\citeauthoryear{Bessi{\`{e}}re}{Bessi{\`{e}}re}{1991}]{Bess91}
{\sc Bessi{\`{e}}re, C.} 1991.
\newblock Arc-consistency in dynamic constraint satisfaction problems.
\newblock In {\em Proceedings of the 9th National Conference on Artificial
  Intelligence, Anaheim, CA, USA, July 14-19, 1991, Volume 1.} 221--226.

\bibitem[\protect\citeauthoryear{Bessiere, Brito, Gutierrez, and
  Meseguer}{Bessiere et~al\mbox{.}}{2014}]{BBGM14}
{\sc Bessiere, C.}, {\sc Brito, I.}, {\sc Gutierrez, P.}, {\sc and} {\sc
  Meseguer, P.} 2014.
\newblock Global constraints in distributed constraint satisfaction and
  optimization.
\newblock {\em Comput. J.\/}~{\em 57,\/}~6, 906--923.

\bibitem[\protect\citeauthoryear{Bessiere, Hebrard, Hnich, Kiziltan, and
  Walsh}{Bessiere et~al\mbox{.}}{2008}]{slide}
{\sc Bessiere, C.}, {\sc Hebrard, E.}, {\sc Hnich, B.}, {\sc Kiziltan, Z.},
  {\sc and} {\sc Walsh, T.} 2008.
\newblock {SLIDE:} {A} useful special case of the {CARDPATH} constraint.
\newblock In {\em {ECAI} 2008 - 18th European Conference on Artificial
  Intelligence, Patras, Greece, July 21-25, 2008, Proceedings}. 475--479.

\bibitem[\protect\citeauthoryear{Bessiere, Katsirelos, Narodytska, Quimper, and
  Walsh}{Bessiere et~al\mbox{.}}{2009}]{decomp}
{\sc Bessiere, C.}, {\sc Katsirelos, G.}, {\sc Narodytska, N.}, {\sc Quimper,
  C.}, {\sc and} {\sc Walsh, T.} 2009.
\newblock Decompositions of all different, global cardinality and related
  constraints.
\newblock In {\em {IJCAI} 2009, Proceedings of the 21st International Joint
  Conference on Artificial Intelligence, Pasadena, California, USA, July 11-17,
  2009}. 419--424.

\bibitem[\protect\citeauthoryear{Bistarelli, Montanari, and Rossi}{Bistarelli
  et~al\mbox{.}}{1997}]{semiring}
{\sc Bistarelli, S.}, {\sc Montanari, U.}, {\sc and} {\sc Rossi, F.} 1997.
\newblock Semiring-based constraint satisfaction and optimization.
\newblock {\em J. {ACM}\/}~{\em 44,\/}~2, 201--236.

\bibitem[\protect\citeauthoryear{Borning, Freeman{-}Benson, and Wilson}{Borning
  et~al\mbox{.}}{1992}]{CH}
{\sc Borning, A.}, {\sc Freeman{-}Benson, B.~N.}, {\sc and} {\sc Wilson, M.}
  1992.
\newblock Constraint hierarchies.
\newblock {\em Lisp and Symbolic Computation\/}~{\em 5,\/}~3, 223--270.

\bibitem[\protect\citeauthoryear{Borning, Maher, Martindale, and
  Wilson}{Borning et~al\mbox{.}}{1989}]{HCLP}
{\sc Borning, A.}, {\sc Maher, M.~J.}, {\sc Martindale, A.}, {\sc and} {\sc
  Wilson, M.} 1989.
\newblock Constraint hierarchies and logic programming.
\newblock In {\em Logic Programming, Proceedings of the Sixth International
  Conference, Lisbon, Portugal, June 19-23, 1989}. 149--164.

\bibitem[\protect\citeauthoryear{Brzozowski, Shallit, and Xu}{Brzozowski
  et~al\mbox{.}}{2009}]{BSX}
{\sc Brzozowski, J.~A.}, {\sc Shallit, J.}, {\sc and} {\sc Xu, Z.} 2009.
\newblock Decision problems for convex languages.
\newblock In {\em Language and Automata Theory and Applications, Third
  International Conference, {LATA} 2009, Tarragona, Spain, April 2-8, 2009.
  Proceedings}. 247--258.

\bibitem[\protect\citeauthoryear{Chandra and Merlin}{Chandra and
  Merlin}{1977}]{CM77}
{\sc Chandra, A.~K.} {\sc and} {\sc Merlin, P.~M.} 1977.
\newblock Optimal implementation of conjunctive queries in relational data
  bases.
\newblock In {\em Proceedings of the 9th Annual {ACM} Symposium on Theory of
  Computing, May 4-6, 1977, Boulder, Colorado, {USA}}. 77--90.

\bibitem[\protect\citeauthoryear{Chatterjee, Doyen, and Henzinger}{Chatterjee
  et~al\mbox{.}}{2010}]{quantlang}
{\sc Chatterjee, K.}, {\sc Doyen, L.}, {\sc and} {\sc Henzinger, T.~A.} 2010.
\newblock Quantitative languages.
\newblock {\em {ACM} Trans. Comput. Log.\/}~{\em 11,\/}~4.

\bibitem[\protect\citeauthoryear{Chaudhuri and Vardi}{Chaudhuri and
  Vardi}{1993}]{ChauVardi}
{\sc Chaudhuri, S.} {\sc and} {\sc Vardi, M.~Y.} 1993.
\newblock Optimization of \emph{Real} conjunctive queries.
\newblock In {\em Proceedings of the Twelfth {ACM} {SIGACT-SIGMOD-SIGART}
  Symposium on Principles of Database Systems, May 25-28, 1993, Washington, DC,
  {USA}}. 59--70.

\bibitem[\protect\citeauthoryear{Colcombet}{Colcombet}{2009}]{Colcombet}
{\sc Colcombet, T.} 2009.
\newblock The theory of stabilisation monoids and regular cost functions.
\newblock In {\em Automata, Languages and Programming, 36th Internatilonal
  Colloquium, {ICALP} 2009, Rhodes, Greece, July 5-12, 2009, Proceedings, Part
  {II}}. 139--150.

\bibitem[\protect\citeauthoryear{Cormen, Leiserson, Rivest, and Stein}{Cormen
  et~al\mbox{.}}{2001}]{CLRS}
{\sc Cormen, T.~H.}, {\sc Leiserson, C.~E.}, {\sc Rivest, R.~L.}, {\sc and}
  {\sc Stein, C.} 2001.
\newblock {\em Introduction to Algorithms, Second Edition}.
\newblock The {MIT} Press and McGraw-Hill Book Company.

\bibitem[\protect\citeauthoryear{Debruyne, Ferrand, Jussien, Lesaint, Ouis, and
  Tessier}{Debruyne et~al\mbox{.}}{2003}]{Debetal03}
{\sc Debruyne, R.}, {\sc Ferrand, G.}, {\sc Jussien, N.}, {\sc Lesaint, W.},
  {\sc Ouis, S.}, {\sc and} {\sc Tessier, A.} 2003.
\newblock Correctness of constraint retraction algorithms.
\newblock In {\em Proceedings of the Sixteenth International Florida Artificial
  Intelligence Research Society Conference, May 12-14, 2003, St. Augustine,
  Florida, {USA}}. 172--176.

\bibitem[\protect\citeauthoryear{Dechter}{Dechter}{2003}]{Dechter}
{\sc Dechter, R.} 2003.
\newblock {\em Constraint processing}.
\newblock Elsevier Morgan Kaufmann.

\bibitem[\protect\citeauthoryear{Dechter and Dechter}{Dechter and
  Dechter}{1988}]{DCSP}
{\sc Dechter, R.} {\sc and} {\sc Dechter, A.} 1988.
\newblock Belief maintenance in dynamic constraint networks.
\newblock In {\em Proceedings of the 7th National Conference on Artificial
  Intelligence. St. Paul, MN, August 21-26, 1988.} 37--42.

\bibitem[\protect\citeauthoryear{Faltings and Macho{-}Gonzalez}{Faltings and
  Macho{-}Gonzalez}{2002}]{openCS}
{\sc Faltings, B.} {\sc and} {\sc Macho{-}Gonzalez, S.} 2002.
\newblock Open constraint satisfaction.
\newblock In {\em Principles and Practice of Constraint Programming - {CP}
  2002, 8th International Conference, {CP} 2002, Ithaca, NY, USA, September
  9-13, 2002, Proceedings}. 356--370.

\bibitem[\protect\citeauthoryear{Faltings and Macho{-}Gonzalez}{Faltings and
  Macho{-}Gonzalez}{2005}]{openCP}
{\sc Faltings, B.} {\sc and} {\sc Macho{-}Gonzalez, S.} 2005.
\newblock Open constraint programming.
\newblock {\em Artif. Intell.\/}~{\em 161,\/}~1-2, 181--208.

\bibitem[\protect\citeauthoryear{Frisch, Hnich, Kiziltan, Miguel, and
  Walsh}{Frisch et~al\mbox{.}}{2002}]{lex}
{\sc Frisch, A.~M.}, {\sc Hnich, B.}, {\sc Kiziltan, Z.}, {\sc Miguel, I.},
  {\sc and} {\sc Walsh, T.} 2002.
\newblock Global constraints for lexicographic orderings.
\newblock In {\em Principles and Practice of Constraint Programming - {CP}
  2002, 8th International Conference, {CP} 2002, Ithaca, NY, USA, September
  9-13, 2002, Proceedings}. 93--108.

\bibitem[\protect\citeauthoryear{Gavanelli, Lamma, Mello, and Milano}{Gavanelli
  et~al\mbox{.}}{2005}]{incomplete_domains}
{\sc Gavanelli, M.}, {\sc Lamma, E.}, {\sc Mello, P.}, {\sc and} {\sc Milano,
  M.} 2005.
\newblock Dealing with incomplete knowledge on clp(\emph{FD}) variable domains.
\newblock {\em {ACM} Trans. Program. Lang. Syst.\/}~{\em 27,\/}~2, 236--263.

\bibitem[\protect\citeauthoryear{Georget, Codognet, and Rossi}{Georget
  et~al\mbox{.}}{1999}]{retraction}
{\sc Georget, Y.}, {\sc Codognet, P.}, {\sc and} {\sc Rossi, F.} 1999.
\newblock Constraint retraction in {CLP(FD):} formal framework and performance
  results.
\newblock {\em Constraints\/}~{\em 4,\/}~1, 5--42.

\bibitem[\protect\citeauthoryear{Gervet}{Gervet}{1997}]{conjunto}
{\sc Gervet, C.} 1997.
\newblock Interval propagation to reason about sets: Definition and
  implementation of a practical language.
\newblock {\em Constraints\/}~{\em 1,\/}~3, 191--244.

\bibitem[\protect\citeauthoryear{He, Flener, and Pearson}{He
  et~al\mbox{.}}{2013}]{badSR}
{\sc He, J.}, {\sc Flener, P.}, {\sc and} {\sc Pearson, J.} 2013.
\newblock Underestimating the cost of a soft constraint is dangerous:
  revisiting the edit-distance based soft regular constraint.
\newblock {\em J. Heuristics\/}~{\em 19,\/}~5, 729--756.

\bibitem[\protect\citeauthoryear{Hentenryck and Provost}{Hentenryck and
  Provost}{1991}]{VHLP}
{\sc Hentenryck, P.~V.} {\sc and} {\sc Provost, T.~L.} 1991.
\newblock Incremental search in constraint logic programming.
\newblock {\em New Generation Comput.\/}~{\em 9,\/}~3/4, 257--276.

\bibitem[\protect\citeauthoryear{Hopcroft and Ullman}{Hopcroft and
  Ullman}{1979}]{HU}
{\sc Hopcroft, J.} {\sc and} {\sc Ullman, J.} 1979.
\newblock {\em Introduction to Automata Theory Languages and Computation}.
\newblock Addison-Wesley.

\bibitem[\protect\citeauthoryear{Ioannidis and Ramakrishnan}{Ioannidis and
  Ramakrishnan}{1995}]{IoannRama}
{\sc Ioannidis, Y.~E.} {\sc and} {\sc Ramakrishnan, R.} 1995.
\newblock Containment of conjunctive queries: Beyond relations as sets.
\newblock {\em {ACM} Trans. Database Syst.\/}~{\em 20,\/}~3, 288--324.

\bibitem[\protect\citeauthoryear{Jaffar and Maher}{Jaffar and
  Maher}{1994}]{JM94}
{\sc Jaffar, J.} {\sc and} {\sc Maher, M.~J.} 1994.
\newblock Constraint logic programming: {A} survey.
\newblock {\em J. Log. Program.\/}~{\em 19/20}, 503--581.

\bibitem[\protect\citeauthoryear{Klug}{Klug}{1988}]{Klug}
{\sc Klug, A.~C.} 1988.
\newblock On conjunctive queries containing inequalities.
\newblock {\em Journal of ACM\/}~{\em 35,\/}~1, 146--160.

\bibitem[\protect\citeauthoryear{Lallouet, Law, Lee, and Siu}{Lallouet
  et~al\mbox{.}}{2011}]{CPstreams}
{\sc Lallouet, A.}, {\sc Law, Y.~C.}, {\sc Lee, J.~H.}, {\sc and} {\sc Siu, C.
  F.~K.} 2011.
\newblock Constraint programming on infinite data streams.
\newblock In {\em {IJCAI} 2011, Proceedings of the 22nd International Joint
  Conference on Artificial Intelligence, Barcelona, Catalonia, Spain, July
  16-22, 2011}. 597--604.

\bibitem[\protect\citeauthoryear{Law and Lee}{Law and Lee}{2004}]{precedence}
{\sc Law, Y.~C.} {\sc and} {\sc Lee, J.~H.} 2004.
\newblock Global constraints for integer and set value precedence.
\newblock In {\em Principles and Practice of Constraint Programming - {CP}
  2004, 10th International Conference, {CP} 2004, Toronto, Canada, September 27
  - October 1, 2004, Proceedings}. 362--376.

\bibitem[\protect\citeauthoryear{Maher}{Maher}{1988}]{equiv}
{\sc Maher, M.~J.} 1988.
\newblock Equivalences of logic programs.
\newblock In {\em Foundations of Deductive Databases and Logic Programming}.
  Morgan Kaufmann, 627--658.

\bibitem[\protect\citeauthoryear{Maher}{Maher}{1993}]{lift}
{\sc Maher, M.~J.} 1993.
\newblock A logic programming view of {CLP}.
\newblock In {\em Logic Programming, Proceedings of the Tenth International
  Conference on Logic Programming, Budapest, Hungary, June 21-25, 1993}.
  737--753.

\bibitem[\protect\citeauthoryear{Maher}{Maher}{2002}]{contig}
{\sc Maher, M.~J.} 2002.
\newblock Analysis of a global contiguity constraint.
\newblock In {\em Proc. Workshop on Rule-Based Constraint Reasoning and
  Programming}.

\bibitem[\protect\citeauthoryear{Maher}{Maher}{2009a}]{synth}
{\sc Maher, M.~J.} 2009a.
\newblock Local consistency for extended {CSP}s.
\newblock {\em Theor. Comput. Sci.\/}~{\em 410,\/}~46, 4769--4783.

\bibitem[\protect\citeauthoryear{Maher}{Maher}{2009b}]{open2}
{\sc Maher, M.~J.} 2009b.
\newblock Open constraints in a boundable world.
\newblock In {\em Integration of {AI} and {OR} Techniques in Constraint
  Programming for Combinatorial Optimization Problems, 6th International
  Conference, {CPAIOR} 2009, Pittsburgh, PA, USA, May 27-31, 2009,
  Proceedings}. 163--177.

\bibitem[\protect\citeauthoryear{Maher}{Maher}{2009c}]{open1}
{\sc Maher, M.~J.} 2009c.
\newblock Open contractible global constraints.
\newblock In {\em {IJCAI} 2009, Proceedings of the 21st International Joint
  Conference on Artificial Intelligence, Pasadena, California, USA, July 11-17,
  2009}. 578--583.

\bibitem[\protect\citeauthoryear{Maher}{Maher}{2009d}]{open3}
{\sc Maher, M.~J.} 2009d.
\newblock {SOGgy} constraints: Soft open global constraints.
\newblock In {\em Principles and Practice of Constraint Programming - {CP}
  2009, 15th International Conference, {CP} 2009, Lisbon, Portugal, September
  20-24, 2009, Proceedings}. 584--591.

\bibitem[\protect\citeauthoryear{Maher}{Maher}{2010}]{open4}
{\sc Maher, M.~J.} 2010.
\newblock Contractibility and contractible approximations of soft global
  constraints.
\newblock In {\em Technical Communications of the 26th International Conference
  on Logic Programming, {ICLP} 2010, July 16-19, 2010, Edinburgh, Scotland,
  {UK}}. 114--123.

\bibitem[\protect\citeauthoryear{Maher, Narodytska, Quimper, and Walsh}{Maher
  et~al\mbox{.}}{2008}]{MNQW}
{\sc Maher, M.~J.}, {\sc Narodytska, N.}, {\sc Quimper, C.}, {\sc and} {\sc
  Walsh, T.} 2008.
\newblock Flow-based propagators for the {SEQUENCE} and related global
  constraints.
\newblock In {\em Principles and Practice of Constraint Programming, 14th
  International Conference, {CP} 2008, Sydney, Australia, September 14-18,
  2008. Proceedings}. 159--174.

\bibitem[\protect\citeauthoryear{Maher and Stuckey}{Maher and
  Stuckey}{1989}]{MS89}
{\sc Maher, M.~J.} {\sc and} {\sc Stuckey, P.~J.} 1989.
\newblock Expanding query power in constraint logic programming languages.
\newblock In {\em Logic Programming, Proceedings of the North American
  Conference 1989, Cleveland, Ohio, USA, October 16-20, 1989. 2 Volumes}.
  20--36.

\bibitem[\protect\citeauthoryear{M{\'{e}}tivier, Boizumault, and
  Loudni}{M{\'{e}}tivier et~al\mbox{.}}{2007}]{MBL07}
{\sc M{\'{e}}tivier, J.}, {\sc Boizumault, P.}, {\sc and} {\sc Loudni, S.}
  2007.
\newblock All different: Softening alldifferent in weighted csps.
\newblock In {\em 19th {IEEE} International Conference on Tools with Artificial
  Intelligence {(ICTAI} 2007), October 29-31, 2007, Patras, Greece, Volume 1}.
  223--230.

\bibitem[\protect\citeauthoryear{M{\'{e}}tivier, Boizumault, and
  Loudni}{M{\'{e}}tivier et~al\mbox{.}}{2009}]{MBL09}
{\sc M{\'{e}}tivier, J.}, {\sc Boizumault, P.}, {\sc and} {\sc Loudni, S.}
  2009.
\newblock Softening gcc and regular with preferences.
\newblock In {\em Proceedings of the 2009 {ACM} Symposium on Applied Computing
  (SAC), Honolulu, Hawaii, USA, March 9-12, 2009}. 1392--1396.

\bibitem[\protect\citeauthoryear{Mittal and Falkenhainer}{Mittal and
  Falkenhainer}{1990}]{condCSP}
{\sc Mittal, S.} {\sc and} {\sc Falkenhainer, B.} 1990.
\newblock Dynamic constraint satisfaction problems.
\newblock In {\em Proceedings of the 8th National Conference on Artificial
  Intelligence. Boston, Massachusetts, July 29 - August 3, 1990, 2 Volumes.}
  25--32.

\bibitem[\protect\citeauthoryear{Nethercote, Stuckey, Becket, Brand, Duck, and
  Tack}{Nethercote et~al\mbox{.}}{2007}]{MZ}
{\sc Nethercote, N.}, {\sc Stuckey, P.~J.}, {\sc Becket, R.}, {\sc Brand, S.},
  {\sc Duck, G.~J.}, {\sc and} {\sc Tack, G.} 2007.
\newblock Minizinc: Towards a standard {CP} modelling language.
\newblock In {\em Principles and Practice of Constraint Programming - {CP}
  2007, 13th International Conference, {CP} 2007, Providence, RI, USA,
  September 23-27, 2007, Proceedings}. 529--543.

\bibitem[\protect\citeauthoryear{Pachet and Roy}{Pachet and Roy}{1999}]{nvalue}
{\sc Pachet, F.} {\sc and} {\sc Roy, P.} 1999.
\newblock Automatic generation of music programs.
\newblock In {\em Principles and Practice of Constraint Programming - CP'99,
  5th International Conference, Alexandria, Virginia, USA, October 11-14, 1999,
  Proceedings}. 331--345.

\bibitem[\protect\citeauthoryear{Pesant}{Pesant}{2004}]{regular}
{\sc Pesant, G.} 2004.
\newblock A regular language membership constraint for finite sequences of
  variables.
\newblock In {\em Principles and Practice of Constraint Programming - {CP}
  2004, 10th International Conference, {CP} 2004, Toronto, Canada, September 27
  - October 1, 2004, Proceedings}. 482--495.

\bibitem[\protect\citeauthoryear{Petit and Poder}{Petit and
  Poder}{2009}]{softcum}
{\sc Petit, T.} {\sc and} {\sc Poder, E.} 2009.
\newblock The soft cumulative constraint.
\newblock {\em CoRR\/}~{\em abs/0907.0939}.

\bibitem[\protect\citeauthoryear{Petit, R{\'{e}}gin, and Bessi{\`{e}}re}{Petit
  et~al\mbox{.}}{2001}]{softcons}
{\sc Petit, T.}, {\sc R{\'{e}}gin, J.}, {\sc and} {\sc Bessi{\`{e}}re, C.}
  2001.
\newblock Specific filtering algorithms for over-constrained problems.
\newblock In {\em Principles and Practice of Constraint Programming - {CP}
  2001, 7th International Conference, {CP} 2001, Paphos, Cyprus, November 26 -
  December 1, 2001, Proceedings}. 451--463.

\bibitem[\protect\citeauthoryear{Quimper and Walsh}{Quimper and
  Walsh}{2006}]{QW}
{\sc Quimper, C.} {\sc and} {\sc Walsh, T.} 2006.
\newblock Global grammar constraints.
\newblock In {\em Principles and Practice of Constraint Programming - {CP}
  2006, 12th International Conference, {CP} 2006, Nantes, France, September
  25-29, 2006, Proceedings}. 751--755.

\bibitem[\protect\citeauthoryear{R{\'{e}}gin}{R{\'{e}}gin}{1994}]{regin}
{\sc R{\'{e}}gin, J.} 1994.
\newblock A filtering algorithm for constraints of difference in csps.
\newblock In {\em Proceedings of the 12th National Conference on Artificial
  Intelligence, Seattle, WA, USA, July 31 - August 4, 1994, Volume 1.}
  362--367.

\bibitem[\protect\citeauthoryear{R{\'{e}}gin}{R{\'{e}}gin}{1996}]{regin96}
{\sc R{\'{e}}gin, J.} 1996.
\newblock Generalized arc consistency for global cardinality constraint.
\newblock In {\em Proceedings of the Thirteenth National Conference on
  Artificial Intelligence and Eighth Innovative Applications of Artificial
  Intelligence Conference, {AAAI} 96, {IAAI} 96, Portland, Oregon, August 4-8,
  1996, Volume 1.} 209--215.

\bibitem[\protect\citeauthoryear{Rossi, van Beek, and Walsh}{Rossi
  et~al\mbox{.}}{2006}]{CPhandbook}
{\sc Rossi, F.}, {\sc van Beek, P.}, {\sc and} {\sc Walsh, T.}, Eds. 2006.
\newblock {\em Handbook of Constraint Programming}. Foundations of Artificial
  Intelligence, vol.~2.
\newblock Elsevier.

\bibitem[\protect\citeauthoryear{Schiex, Fargier, and Verfaillie}{Schiex
  et~al\mbox{.}}{1995}]{VCSP}
{\sc Schiex, T.}, {\sc Fargier, H.}, {\sc and} {\sc Verfaillie, G.} 1995.
\newblock Valued constraint satisfaction problems: Hard and easy problems.
\newblock In {\em Proceedings of the Fourteenth International Joint Conference
  on Artificial Intelligence, {IJCAI} 95, Montr{\'{e}}al Qu{\'{e}}bec, Canada,
  August 20-25 1995, 2 Volumes}. 631--639.

\bibitem[\protect\citeauthoryear{Schulte and Tack}{Schulte and Tack}{2009}]{ST}
{\sc Schulte, C.} {\sc and} {\sc Tack, G.} 2009.
\newblock Weakly monotonic propagators.
\newblock In {\em Principles and Practice of Constraint Programming - {CP}
  2009, 15th International Conference, {CP} 2009, Lisbon, Portugal, September
  20-24, 2009, Proceedings}. 723--730.

\bibitem[\protect\citeauthoryear{Sellmann}{Sellmann}{2006}]{Sellmann}
{\sc Sellmann, M.} 2006.
\newblock The theory of grammar constraints.
\newblock In {\em Principles and Practice of Constraint Programming - {CP}
  2006, 12th International Conference, {CP} 2006, Nantes, France, September
  25-29, 2006, Proceedings}. 530--544.

\bibitem[\protect\citeauthoryear{van Hoeve, Pesant, and Rousseau}{van Hoeve
  et~al\mbox{.}}{2006}]{warming}
{\sc van Hoeve, W.~J.}, {\sc Pesant, G.}, {\sc and} {\sc Rousseau, L.} 2006.
\newblock On global warming: Flow-based soft global constraints.
\newblock {\em J. Heuristics\/}~{\em 12,\/}~4-5, 347--373.

\bibitem[\protect\citeauthoryear{van Hoeve and R{\'{e}}gin}{van Hoeve and
  R{\'{e}}gin}{2006}]{open}
{\sc van Hoeve, W.~J.} {\sc and} {\sc R{\'{e}}gin, J.} 2006.
\newblock Open constraints in a closed world.
\newblock In {\em Integration of {AI} and {OR} Techniques in Constraint
  Programming for Combinatorial Optimization Problems, Third International
  Conference, {CPAIOR} 2006, Cork, Ireland, May 31 - June 2, 2006,
  Proceedings}. 244--257.

\bibitem[\protect\citeauthoryear{Verfaillie and Jussien}{Verfaillie and
  Jussien}{2005}]{dynCSP}
{\sc Verfaillie, G.} {\sc and} {\sc Jussien, N.} 2005.
\newblock Constraint solving in uncertain and dynamic environments: {A} survey.
\newblock {\em Constraints\/}~{\em 10,\/}~3, 253--281.

\end{thebibliography}

\end{document}